\documentclass[a4paper,12pt]{article}

\usepackage{bbold}
\usepackage{tabu}
\usepackage{graphicx,subcaption}
\usepackage{amsfonts}
\usepackage{amsmath}
\usepackage{amssymb}
\usepackage{tensor}
\usepackage{yhmath}
\usepackage{epsfig}
\usepackage{slashed}
\usepackage{pdfpages}
\usepackage{graphicx,epstopdf}
\colorlet{darkblue}{blue!70!black}
\newcommand{\arxiv}[1]{\href{http://www.arXiv.org/abs/#1}{#1}}
\usepackage[colorlinks=true,urlcolor=darkblue,linktocpage=true,linkcolor=darkblue,citecolor=blue]{hyperref}
\usepackage{array}
\usepackage{tikz}
\usepackage{mathtools}
\usepackage{braket}
\usepackage{enumitem}
\usepackage{bm}
\usepackage[thinlines]{easytable}
\usepackage{cancel}
\usepackage{leftidx}
\usepackage{cite}
\usepackage{cancel}
\usepackage[normalem]{ulem}
\numberwithin{equation}{section}

\newcommand{\mO}{\mathcal{O}}

\newcommand{\eps}{\epsilon}

\newcommand{\no}{\nonumber}
\newcommand{\der}{\partial}
\newcommand{\be}{\begin{equation}}
\newcommand{\ee}{\end{equation}}

\graphicspath{{./figures/}}
\newcommand{\Tr}{{\rm Tr}}
\newcommand{\del}{\partial}
\newcommand{\bea}{\begin{eqnarray}}
\newcommand{\eea}{\end{eqnarray}}
\newcommand{\ba}{\begin{eqnarray}}
\newcommand{\ea}{\end{eqnarray}}

\def\ie{{\it i.e.}}
\def\eg{{\it e.g.}}

\def\a{\alpha}
\def\b{\beta}

\def\eps{\epsilon}

\def\t{\tilde}

\def\O{{\cal O}}

\def\F{{\cal F}}

\def\({\left (}
\def\){\right )}
\def\[ {\left[}
\def\]{\right]}

\def\bz{{\bar z}}

\def\bw{{\bar w}}
\def\bT{{\bar T}}
\def\tE{{{t_{E}}}}
\def\arcsinh{{\rm arcsinh}\, }

\begin{document}


\title{Boundary-induced transitions \\ in M\"obius quenches of holographic BCFT}

\author{Alice Bernamonti$^{\star,\ast}$, Federico Galli$^{\ast}$, Dongsheng Ge$^{\dagger}$}
\date{}
 %

\maketitle

\vspace*{-7.3cm}
\begin{flushright}
OU-HET-1222\\
\end{flushright}
\vspace*{0.2cm}

\vspace*{5cm}

\centerline{\it $^{\star}$Dipartimento di Fisica e Astronomia, Universit\'a di Firenze;}
\centerline{\it Via G. Sansone 1;  I-50019 Sesto Fiorentino, Italy}
\centerline{\it $^\ast$INFN, Sezione di Firenze;}
\centerline{\it Via G. Sansone 1; I-50019 Sesto Fiorentino, Italy}
\centerline{ \it $\,^\dagger$Department of Physics, Osaka University,}
\centerline{\it Machikaneyama-Cho 1-1, Toyonaka 560-0043, Japan}

\vskip 4cm 
 
\begin{abstract}
 
Boundary effects play an interesting role in finite-size physical systems.  
In this work, we study the boundary-induced properties of 1+1-dimensional critical systems driven by inhomogeneous M\"obius-like quenches. We focus on the entanglement entropy in BCFTs with a large central charge and a sparse spectrum of low-dimensional operators. We find that the choice of boundary conditions leads to different scenarios of dynamical phase transitions.  
We also derive these results in a holographic description in terms of intersecting branes in AdS$_3$, and find a precise match.
\end{abstract}

\vskip 1.5cm 

\thispagestyle{empty}

\newpage

\tableofcontents


\section{Introduction}

Ordinary physical systems have boundaries. These introduce boundary effects, which can be probed by physical quantities, such as the partition  and correlation functions.  Of particular interest are conformally invariant systems with conformal boundaries, described by boundary conformal field theory (BCFT). In 1+1 dimensions this subject was mainly laid out by Cardy \cite{Cardy:1989ir,Cardy:1991tv,Cardy:2004hm}, and it is the main focus of this work. 
 
In 2d BCFT, boundary effects are captured by the boundary entropy $s_{\mathcal B} = \ln g_{\mathcal B} \equiv  \ln \langle 0 \ket{\mathcal B}$, which measures the ground state degeneracy for a given conformal boundary condition $\mathcal B$ \cite{Affleck:1991tk,Calabrese:2004eu,Calabrese:2009qy}. This can be evaluated explicitly in rational conformal field theories once the solutions to Cardy's equation are found \cite{Cardy:1989ir,Behrend:1998fd,Behrend:1999bn}. For an infinite strip of width $L$, the boundary conditions for its two edges can be chosen differently. 
As a result, the distinct boundary conditions lead to two different boundary entropies. In rational CFT they can assume a limited number of values due to the presence of a finite number of boundary primaries \cite{Behrend:1999bn}. 

Here we are concerned mainly with holographic CFTs, which have a large central charge $c$ and a sparse spectrum of  light operators \cite{Heemskerk:2009pn,Hartman:2014oaa}. Large part of the recent work in holographic BCFT is connected with the developments in understanding entanglement in gravitational systems, via the entanglement island proposal \cite{Penington:2019npb,Almheiri:2019psf,Almheiri:2019hni,Almheiri:2019qdq,Almheiri:2020cfm}. This has sparked renewed interest in refining the understanding of the AdS/BCFT correspondence \cite{Takayanagi:2011zk,Fujita:2011fp}.  
In the simplest AdS/BCFT models, boundary conditions are implemented via the insertion of end of the world (EOW) branes anchored on the spacetime boundary and extending into the AdS bulk \cite{Takayanagi:2011zk,Fujita:2011fp}. A number of features emerge for holographic BCFTs realized in this way \cite{Takayanagi:2011zk,Fujita:2011fp,Rozali:2019day,Sully:2020pza,Geng:2021iyq,Reeves:2021sab,Miyaji:2021ktr,Biswas:2022xfw,Miyaji:2022dna,Kanda:2023zse,Suzuki:2022tan,Kanda:2023jyi}. In particular, they  have a continuous spectrum of heavy operators, with conformal weight proportional to the central charge, up to the black hole threshold $c/24$. Also, the tension $T_i$ of the EOW  branes in the AdS bulk determines the boundary entropy $s_i =  \frac{c}{6} \text{arctanh}\, T_i$ for the dual boundary state  \cite{Takayanagi:2011zk,Fujita:2011fp}. 
The brane tension can take values continuously in the range $|T_i| \le 1$ and thus yields for these models a  boundary entropy taking arbitrary values and scaling with the CFT central charge. 
Indeed, in the holographic computation of the entanglement entropy in AdS/BCFT models, the boundary entropy arises from the Ryu-Takayanagi formula applied to bulk geometries bounded by EOW branes \cite{Takayanagi:2011zk,Fujita:2011fp}. 

In holographic BCFT, different competing phases in the evaluation of the entanglement entropy may arise, and the boundary entropy plays a role in determining the dominant one. These situations include holographic computations in terms of the Ryu-Takayanagi surfaces \cite{Takayanagi:2011zk,Fujita:2011fp},   but also first principles CFT calculations in specific limits, and applications to double holographic models via the entanglement island formula as in, \eg, \cite{Caputa:2019avh,Sully:2020pza,Geng:2021iyq,Bianchi:2022ulu,Penington:2019npb,Almheiri:2019psf,Almheiri:2019hni,Almheiri:2019qdq}.

In far-from-equilibrium settings, such as those generated by quantum quenches, the system may experience dynamical transitions between these competing phases. In this work, we consider spatially inhomogeneous quenches in 2d BCFT. In particular, we focus on a family of processes in which the Hamiltonian is abruptly changed to the so called M\"obius Hamiltonian \cite{Gendiar:2008udd,Hikihara:2011mtb,Katsura:2011ss,Gendiar:2010,Okunishi:2016zat,Wen:2016inm,Goto:2021sqx,Goto:2023wai,Nozaki:2023fkx}. A well-studied limit of this class is known as sine-square-deformed (SSD) Hamiltonian, which was originally introduced to suppress boundary effects and efficiently isolate bulk properties \cite{Gendiar:2008udd,Gendiar:2010,Hikihara:2011mtb,Shibata:2011jup,Maruyama:2011njv,Katsura:2011zyx,Katsura:2011ss}. In fact, it was observed that in 1+1-dimensional CFT the SSD Hamiltonian shares the same ground state of a uniform system with periodic boundary conditions and that, to a certain extent, the system size is effectively infinite in this limit \cite{Okunishi:2016zat,Wen:2016inm,Tada:2014kza,Ishibashi:2015jba,Ishibashi:2016bey}. 

As we summarize below, here we deal with an instance that combines the above ingredients. For a BCFT on an interval with mixed boundary conditions, we study the entanglement entropy evolution following an inhomogeneous quench by a M\"obius-like deformation.  
We will show that two competing phases in the entanglement entropy arise from the presence of mixed boundary conditions, and that a dynamical transition from one phase to the other is controlled by the quench protocol. This is a new feature as compared to previous studies on this class of quenches, where boundary effects were found to be suppressed \cite{Gendiar:2008udd,Hikihara:2011mtb,Katsura:2011ss,Gendiar:2010,Okunishi:2016zat}. 

\paragraph{Summary \\}

\begin{figure}[t]
\centering
        \includegraphics[width=.9\textwidth]{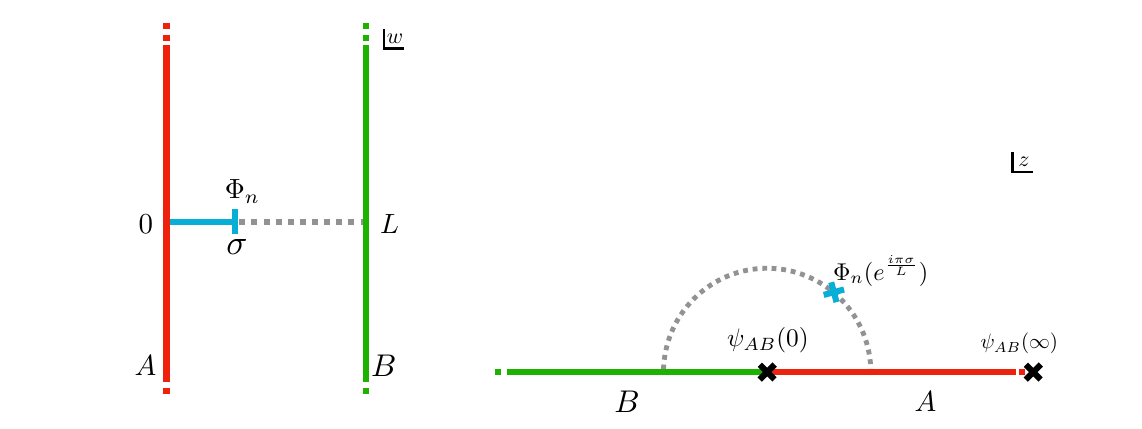}
\caption{Left: The physical system is defined on a strip of spatial width $L$, with conformal boundary conditions $A$ and $B$ a the two edges. We consider a bipartition of the system defined considering a spatial interval of size $\sigma$ adjacent to the boundary with conformal boundary condition $A$. Via the replica trick the evaluation of the entanglement entropy for this bipartition is mapped to calculating the one point fucntion of a replica twist operator $\Phi_n$ inserted at the endpoint of the interval. Right: The system mapped to the upper half plane. The change in boundary conditions is sustained by the insertion of a boundary condition changing operator $\psi_{AB}$.}
\label{fig:Setup}
\end{figure}

The dynamical process we consider is  the following. We prepare a state $\ket{\psi_{AB}}$ on a finite-size system of length $L$, with distinct boundary conditions $A$ and $B$, as illustrated in fig. \ref{fig:Setup}. 
This is achieved inserting boundary condition changing (BCC) operators, which mediate the change in boundary conditions \cite{Affleck:1996mm}. BCC operators are not  local operators of the boundary CFT in the conventional sense, but are treated as boundary primaries with similar scaling properties as bulk primaries  \cite{Cardy:2004hm}. In particular  $\ket{\psi_{AB}}$  can be seen as an eigenstate of the BCFT Hamiltonian.  
 
At $t=0$ the state is quenched by  turning on a deformation of the BCFT Hamiltonian. For $t>0$ the system is driven with such a deformed Hamiltonian and undergoes a non-trivial evolution.\footnote{In spirit, this is the same thing happening in the prototypical global quench protocol of Calabrese and Cardy \cite{Calabrese:2005in}. In both cases the initial state is not an eigenstate of the Hamiltonian that drives the dynamics for $t>0$.}  We analyse the dynamics following from different deformations: the SSD, M\"obius  and generalized deformation of section  \ref{sec:GeneralDeform}. A key technical advantage is that these allow to explicitly obtain the time evolution in a closed form \cite{Okunishi:2016zat,Wen:2016inm,Wen:2018vux}.

As a probe of the dynamics, we consider the real-time evolution of the entanglement entropy for an interval adjacent to one end of the system. 
Via the replica trick, the computation of the entanglement entropy is recast in evaluating a three-point function of two BCC operators and a replica twist in the bulk. 
The details of such a correlator depend on the specific theory. We extract and analyse the contribution coming from the conformal family of the stress tensor.  
The reason is twofold. On the one hand, this contribution is universal. It is present in every 2d CFT with the identity operator in its normalizable spectrum, and it does not depend on CFT data other than the central charge and the ground state degeneracy. On the other hand, in holographic CFTs the contribution of the identity conformal family is expected to dominate, and to match dual holographic calculations in three dimensional semiclassical  gravity. 

With mixed boundary conditions the approximate result obtained from the conformal family of the identity operator is not unique. There are different boundary operator expansions of a bulk CFT operator one can consider \cite{Lewellen:1991tb}. In particular, one associated with boundary condition $A$ and one with boundary condition $B$. These give competing channels in the evaluation of the correlator. 
Already in the initial state  what channel, or phase, dominates depends on the size of the interval considered and difference in boundary entropies $s_{A,B}$ \cite{Geng:2021iyq,Biswas:2022xfw,Miyaji:2022dna}.  More interestingly, we show that the quench dynamics can drive a transition from one phase to the other (and back) as time evolves. 

We find that the dynamical phase transition pattern is determined by the relative amount of  boundary degrees of freedom associated to the two boundaries. Depending on the difference between $s_A$ and $s_B$, the entanglement entropy can exhibit or not a transition between phase $A$ and phase $B$. Under the SSD dynamics the transition can only happen once, while it becomes recurrent for the M\"obius-like quenches, with characteristic timescales determined by the details of the deformation. 

This finds an exact match and a simple interpretation in the holographic gravitational bulk description obtained in terms of  AdS$_3$ bounded by two intersecting  EOW branes \cite{Miyaji:2022dna,Biswas:2022xfw}. We show that the entanglement entropy obtained in the BCFT computation for the two phases is in one to one correspondence with the length of minimal geodesics. Indeed for a point on the AdS$_3$ boundary, there exists in general two competing Ryu-Takayanagi geodesics, each ending on one of the two EOW branes. We also show that the entanglement dynamics can be matched exploiting the local equivalence of AdS$_3$ geometries and  extending the map relating the initial and time evolved state in the BCFT to a diffeomorphism in the bulk.  \\

The rest of the paper is organized as follows, in section \ref{sec:strip},  we introduce the setup and relevant BCFT technicalities needed for our analysis. In particular, we show how the initial state of our system is defined and compute the corresponding entanglement entropy. 
In section \ref{sec:EESSDM}, we introduce the SSD and M\"obius Hamiltonian on the strip, and analyze the time evolution of the entanglement entropy for a quench with these Hamiltonians. 
In section  \ref{sec:GeneralDeform},  we consider the generalization of this analysis  to the case of deformed Hamiltonians  obtained from   $sl(2,\mathbb{R})$  subalgebras of Virasoro generators at arbitrary levels.  Section \ref{sec:holography} describes the holographic dual of our setup and analysis. Using a bulk description in terms of AdS$_3$ with intersecting EOW branes, we find an exact match of the BCFT computation with the holographic result. We conclude in section \ref{sec:conclusions}, where we comment on the broader applicability of our analysis and results, and give future perspectives.

\section{BCFT with mixed boundary conditions} \label{sec:strip}
 
In this section we review how to characterize a CFT on an interval with different conformal boundary conditions, $A$ and $B$, at the two ends   \cite{Lewellen:1991tb,Affleck:1996mm,Cardy:2004hm}. The state with mixed boundary conditions we will consider, $| \psi_{AB} \rangle $, will either be the ground state or a high energy eigenstate  for the BCFT with distinct boundary conditions. 

We will review how to evaluate the entanglement entropy in this state for an interval adjacent to one of the boundaries. Via the replica approach, this reduces to evaluating a replica twist one-point  function on the strip with mixed boundary conditions. Specializing to the case of holographic CFTs, we will approximate the correlator with the universal contribution of the stress tensor and its conformal family. This is motivated by the assumptions in an holographic CFT on the sparseness of the spectrum of low dimensional operators and on the dominance of the identity Virasoro conformal block (see, \eg, \cite{Heemskerk:2009pn,Hartman:2013mia,Fitzpatrick:2014vua,Fitzpatrick:2015zha,Caputa:2014eta,Hartman:2014oaa,Asplund:2014coa,Perlmutter:2015iya,Anous:2016kss,Balasubramanian:2017fan} for examples and discussions).  Let us remark that this is a working assumption and here we are not proving the validity of this approximation. We will however show explicitly in section \ref{sec:holography} that the result obtained in this way exactly captures the result following from the corresponding dual  computation in terms of Ryu-Takayanagi surfaces \cite{Ryu:2006bv,Takayanagi:2011zk,Fujita:2011fp,Hubeny:2007xt}.

\subsection{Strip with mixed boundary conditions}\label{sec:BCCop}

We consider a state in a CFT on an interval with two different conformally invariant boundary conditions. In particular, in the Euclidean plane $w = \tau + i \sigma$, we consider an infinite strip $-\infty <  \tau < \infty$ of width $0 \le \sigma \le L$. Two conformally invariant boundary conditions $A,B$ are imposed along $\sigma =0$ and $\sigma =L$ respectively. With  
\be \label{eq:toUHP}
z =  e^{\frac{\pi}{L} w} 
\ee
the strip is mapped into the upper half plane (UHP), with the two boundaries $\sigma =0, L$  mapped respectively to the positive and negative real axis $z = x$.  In the UHP  the change in conformal boundary conditions along the boundary at $x =0, \infty$ is mediated by boundary condition changing (BCC) operators $\psi_{AB}(x)$  (see fig.~\ref{fig:Setup}) \cite{Lewellen:1991tb,Cardy:2004hm}. 
More precisely, starting in the UHP with a state with homogeneous boundary condition $A$, the mixed boundary conditions $A$ and $B$ state is obtained as\footnote{For notational simplicity and since it is clear from the context, we do not distinguish between the BCC operators  $\psi_{AB}$ and $ \psi_{BA}$ \cite{Lewellen:1991tb}.}
\be \label{eq:ABstate}
\ket{\psi_{AB}} = \psi_{AB} (0) \ket{0}_{AA}  \qquad 
 \bra{\psi_{AB}} =   \lim_{x\to \infty} \frac{x^{2h_\psi}}{c_{\psi \psi}}\,  {}_{AA}\bra{0} \psi_{AB}(x)  \equiv  {}_{AA}\bra{0}  \psi_{AB}(\infty)
 \,. 
\ee
 BCC operators act non-locally to implement the change of boundary conditions, but can be treated as primaries with weight $h_\psi$ in the CFT defined on the boundary \cite{Lewellen:1991tb,Cardy:2004hm}.   
  $c_{\psi \psi}$ is   the coefficient  of the two-point function of the BCC operator, so that the state is normalized $ \bra{ \psi_{AB}}\psi_{AB} \rangle =1$. The conformal dimension $h_\psi$  is determined in terms of gap in energy between the initial state, with homogeneous boundary conditions $A$, and the state with mixed boundary conditions $A$ and $B$. In particular  \cite{Affleck:1996mm}
\be \label{hBCC}
h_\psi = \frac{L \, \Delta E}{\pi}  = \frac{L}{\pi} \( E_{AB} - E_{AA}\) \,, 
\ee
where $E_{AA}$ is the energy of the ground state on the strip with boundary conditions $A$. $E_{AB}$ instead is the energy of the lowest energy eigenstate on the strip with mixed boundary conditions with a non-zero overlap with $\ket{\psi_{AB}}$. Notice that depending on the operator $\psi_{AB}$, this may correspond to the ground state or an excited state of the BCFT with mixed boundary conditions \cite{Affleck:1996mm}.

We are interested in holographic BCFTs, which have a dual description in terms of  semiclassical gravitational configurations in AdS$_3$. For these, the change in boundary conditions holographically translates into a change of the asymptotically  AdS$_3$ geometry  \cite{Takayanagi:2011zk,Fujita:2011fp,Akal:2020wfl}. The corresponding change in energy is thus proportional to the central charge $c$. We will then be concerned with the case where the BCC has a conformal weight proportional to the central charge $c$ of the BCFT \cite{Geng:2021iyq,Biswas:2022xfw,Miyaji:2022dna}. As we will review in section~\ref{sec:holography}, in bottom-up models of holographic CFTs the BCC conformal operators have a continuous spectrum in the range  $0<h_\psi/c \leq 1/24$ \cite{Biswas:2022xfw,Miyaji:2022dna}. When talking of holographic BCFT  in the rest of this work we will then assume implicitly this range of values. 

 \subsection{Twist operator correlator}

We want to evaluate the entanglement entropy for a spatial region consisting of a portion of the strip that includes one of the boundaries. Without loss of generality, we pick the entangling region to be a portion of dimension  $0<\sigma<L$ of the entire system, adjacent to the origin. The boundary included in this region is therefore the one with boundary condition $A$, see fig.~\ref{fig:Setup}.

Using the replica trick, the entanglement entropy $S$ can be obtained in terms of the R\'enyi entropies $S^{(n)}_\sigma$ (for a review see \cite{Calabrese:2009qy}). For any integer $n\geq 2$, these are defined from the reduced density matrix $\rho_\sigma$ associated to the geometric region we are considering as
\be
S^{(n)}_\sigma = \frac{1}{1-n}\log \Tr \rho_\sigma^n \, .
\ee
The computation of $S^{(n)}_\sigma$ can be recast into the evaluation of a correlator of twist operators in the strip geometry with mixed boundary  conditions. For the case at hand this is a one-point  function\footnote{Here are considering a time independent state, so we are setting  $\tau=0$ without loss of generality.}
\be \label{eq:Sn}
S^{(n)}_\sigma = \frac{1}{1-n} \log \bra{\psi_{AB}} \Phi_n(w= i \sigma,\bar w=- i\sigma)\ket{\psi_{AB}}  \, , 
\ee
with the twist operator $\Phi_n$ inserted at the endpoint of the entangling interval (see fig.~\ref{fig:Setup}).
The entanglement entropy $S$ is then obtained as the analytic continuation to  $n = 1$. Schematically,
\be  \label{eq:EE}
S = \lim_{n \to 1} S^{(n)}_\sigma =  \lim_{n \to 1}\frac{1}{1-n} \log \bra{\psi_{AB}} \Phi_n(w,\bar w)  \ket{\psi_{AB}} \, .
\ee

The above correlator is not defined in the original CFT, but in the cyclic orbifold theory  CFT$^n/\mathbb Z_n$ resulting from using the replica trick.\footnote{In defining the state $\ket{\psi_{AB}}$  in \eqref{eq:Sn} one would actually   need to consider  the  operator $\Psi_{AB}=\psi_{AB}^{\otimes n}$ in the orbifold CFT$^n/\mathbb Z_n$, instead of $\psi_{AB}$. Nonetheless, for all practical purposes,  in the limit    \eqref{eq:EE}  one can simply keep  defining the state  with $\psi_{AB}$. Any appearance of $\Psi_{AB}$ via $h_{\Psi} = n h_\psi$ would  reduce to the effect of having  $\psi_{AB}$ with  $h_\psi$. 
See, \eg, \cite{Asplund:2014coa, Bianchi:2022ulu} for similar computations.} 
The corresponding replica twist field has conformal dimension
\be \label{eq:htwist}
h_h = \bar h_n = \frac{c}{24}\(n - \frac{1}{n}\) \, .
\ee

The strip correlator on the right hand side of \eqref{eq:EE} is   conveniently  evaluated via a map to the UHP. Using    $z =  e^{\frac{\pi}{L} w}$ and owing to the normalization of the state we have\footnote{The UHP expectation value is homogeneous with uniform boundary condition $A$, that is $\langle  \dots  \rangle_{\text{UHP}}   =   {}_{AA}\bra{0}   \dots \ket{0}_{AA}$.}
\bea
\bra{\psi_{AB}} \Phi_n(w,\bar w)  \ket{\psi_{AB}}
= \( \frac{\del z}{\del w} \)^{h_n}   \( \frac{\del \bz}{\del \bw} \)^{h_n}  \langle \psi_{AB}(0)  \Phi_n (z,\bar z)  \psi_{AB} (\infty) \rangle_{\text{UHP}} \, .  \label{eq:3ptUHP}
\eea

This three point function in the UHP involves two boundary and one bulk operator and is thus not completely fixed by conformal invariance. Similarly to a four-point function  in the full complex plane, it can be expanded in the sum of conformal Virasoro blocks. 
There are in fact two different expansions, or channels, one can consider here  \cite{Lewellen:1991tb}. One can perform the boundary expansion of a primary bulk operator $\Phi$, in our case a  bulk twist operator, assuming it approaches the boundary where the boundary condition is fixed to be  $A$.  Alternatively, $\Phi$ can expanded in terms of boundary operators assuming the condition on the boundary is $B$.  The two corresponding conformal block expansions can be written as  \cite{Lewellen:1991tb}
\be \label{eq:Achann}
\langle \psi_{AB}(0)  \Phi (z,\bar z)  \psi_{AB}(\infty) \rangle_{\text{UHP}} =  (z - \bz)^{-2h_\Phi} (1- \eta )^{2h_\Phi}  \sum_p C^{A}_{p}   \F_p(1-\eta)\, ,
\ee 
and
\be \label{eq:Bchann}
\langle\psi_{AB} (0)  \Phi (z,\bar z)\psi_{AB}(\infty) \rangle_{\text{UHP}} = ( z - \bz)^{-2h_\Phi} (1- \eta )^{2h_\Phi}  \sum_p C^{B}_{p}   \F_p(1-   e^{-2\pi i }\eta)\, 
\ee 
where we defined the ratio 
\be
\eta = \frac{z}{\bar z} \, . 
\ee
In the above expressions, the  sum is over exchanged primaries of the CFT defined on the boundary, with  $C^A_p$ and $C^B_p$ constants related to   operator  expansion coefficients (see  \cite{Lewellen:1991tb} for details).
$\F_p$  is the same function defining the  holomorphic  Virasoro conformal block in the full complex plane.  More precisely, take in the full complex plane a primary $\phi$ having the same conformal weight as $\Phi$ and a primary $\psi$ with the same conformal weight as $\psi_{AB}$. $\F_p(1-\eta)$  is then the holomorphic Virasoro conformal block appearing in the t-channel expansion  
\be \label{eq:tchann}
\langle \psi(0) \phi(\eta, \bar \eta) \phi(1)  \psi(\infty) \rangle_{\text{plane}}=\sum_p a_{p} \F_p(1-\eta)\bar\F_p(1-\bar \eta) \, . 
\ee 
Notice that despite the full correlator being single valued in the complex plane,  $\F_p$ is in general multivalued  in the $\eta$ plane and requires a choice of branch cut.   Here we are   taking the branch cut of $\F_p(1-\eta)$  in such a way as to recover the expected short distance singularity in the OPE limit $\phi(\eta, \bar \eta) \to \phi(1)$. Having fixed that, the explicit phase in front of $\eta$ in  \eqref{eq:Bchann} differentiates the single terms in the expansion from those in \eqref{eq:Achann}. Of course, the exact correlator obtained from the full sums is single valued. 

As discussed in   section \ref{sec:BCCop}, we will consider the case of heavy BCC operators $\psi_{AB}$ with conformal weight  $h_\psi$ scaling with the central charge $c$.  In our computation the bulk operator above is identified with a replica twist $\Phi_n$. Its weight $h_n$ also scales with the central charge of the CFT, but $h_n /c \to 0$ in the limit  $n \to 1$.  In the holographic CFT literature this is referred to as a (perturbatively) light operator (see, \eg, \cite{Fitzpatrick:2014vua,Fitzpatrick:2015zha}). 
The relevant Virasoro conformal blocks $\F_p$  are therefore those for an  Heavy-Heavy-Light-Light four point function \eqref{eq:tchann}. At the leading order in a large $c$ expansion these are known  in a closed form \cite{Fitzpatrick:2014vua,Fitzpatrick:2015zha}.

\subsection{Universal contribution from the stress tensor } \label{sec:univSEE}

The  two expansions  \eqref{eq:Achann} and  \eqref{eq:Bchann}   give completely equivalent results when evaluating  the full correlator $\langle \psi(0)  \Phi_n(z,\bar z)  \psi (\infty) \rangle_{\text{UHP}}$.   However, for holographic CFTs the full CFT data are not known  and we can only provide an approximate result. 
We will consider the universal contribution of the stress tensor coming from the identity Virasoro block and assume that in an holographic CFT the correlator is well approximated by this contribution alone (see, \eg, \cite{Hartman:2013mia,Fitzpatrick:2014vua,Fitzpatrick:2015zha,Caputa:2014eta,Asplund:2014coa,Anous:2016kss,Balasubramanian:2017fan} for work in similar context and more detailed discussions of this approximation). 

Roughly speaking, this approximation is motivated by the fact that for holographic CFTs the sum over Virasoro conformal blocks is a sum of exponentials, weighted by the conformal dimension of the exchanged operator \cite{Zamolodchikov:1984eqp,Zamolodchikov:1987}.  Applying a saddle-point approach, the sum is then approximated by the largest term. For holographic CFTs, which have a sparse spectrum of low-dimensional operators, this is given by the identity Virasoro block. In making this approximation one is also assuming that other contributions from light operators are suppressed in the large $c$ expansion, and ignoring possible non-perturbative terms. In addition, the universal contribution of the stress energy tensor alone is expected  to encode the geometric features of the dual holographic description, as we will show in section \ref{sec:holography}.

The different expansions channels in \eqref{eq:Achann} and  \eqref{eq:Bchann} provide competing identity Virasoro block approximations.   Following the saddle point logic, one is then lead to consider these channels as competing, and to retain the dominant saddle point contribution. 
Here we  shall notice that  we expect each of the identity Virasoro contributions to provide a good approximation to the full correlator only for some range of the parameters. When knowing the full correlator one would observe a smooth transition connecting these regimes \cite{Headrick:2010zt}. In our working approach, where we only retain two competing saddles provided by the identity Virasoro block, we will instead observe a sharp transition.
 
To  evaluate the correlator, we then  consider the explicit expression for  the Virasoro identity block at leading order in a large $c$ expansion  \cite{Fitzpatrick:2014vua,Fitzpatrick:2015zha}   
\be \label{eq:idblock}
\F_0 (1-\eta)= \eta^{(\alpha-1) h_n } \(\frac{1 - \eta^{\alpha}}{\alpha} \)^{-2h_n }  \, .
\ee 
Here
\be \label{eq:alphadef}
\alpha  = \sqrt{1 - \frac{24 h_\psi }{c} }\, ,
\ee 
and $h_n$ is the conformal weight \eqref{eq:htwist} of the twist operator $\Phi_n$. Notice that  \eqref{eq:idblock} is  only valid  for $h_n/c \ll1$, \ie, for $n$ close to one.  

At leading order in the $\eta \to 1$ limit
\be
\F_0 (1-\eta)  \approx (1- \eta)^{-2h_n} \, .
\ee 
Taking this limit in  \eqref{eq:Achann} selects the identity contribution in the bulk to boundary OPE of the bulk operator $\Phi_n$. This 
shows  how the coefficient $C^{A}_{0}$ in \eqref{eq:Achann} is  related to the expectation value of $\Phi_n$ in the UHP with homogeneous boundary conditions $A$.  Explicitly  
\be \label{eq:CA0}
\langle \Phi_n(z,\bz)\rangle_{\text {UHP}}= \frac{A_{\Phi_n}}{|z - \bz|^{2h_n}} = \frac{C^A_0}{(z - \bz)^{2h_n}} \, ,
\ee
where we are neglecting the regularization associated to the insertion of a twist operator,  which will be restored in the final result for the entanglement entropy. 

Approximating the full three-point function with the identity contribution  in the $A$-channel  \eqref{eq:Achann} then gives 
\be
\begin{aligned} \label{eq:AUHP}
\bra{\psi_{AB} }\Phi_n (z, \bz)   \ket{\psi_{AB}}_{\text {UHP}}&\approx C_0^A \alpha^{2h_n} (z \bz)^{(\alpha-1) h_n} ( z^\alpha - \bz^\alpha)^{-2h_n} \\
&= C_0^A\alpha^{2h_n} (z \bz)^{- h_n } \[ 2 \sinh \frac{\alpha}{2} \log \frac{z}{\bz} \]^{-2h_n} \,,
\end{aligned}
\ee
which via \eqref{eq:3ptUHP} in terms of the physical strip setup yields
\be \label{eq:A3pt}
\bra{\psi_{AB} }\Phi_n (w, \bw)   \ket{\psi_{AB}}  \approx  
  A_{\Phi_n}  \( \frac{\alpha \pi }{2 L}\)^{2h_n} \left[  \sin  \alpha \pi   \frac{ \sigma }{ L}  \right]^{- 2h_n }  \, .
\ee  
In the last expression we have used the fact that  the twist  insertion is at $w = i \sigma $. In the limit $\sigma\to0$ this result reproduces  the  result for a small entangling region adjacent to the boundary at $\sigma=0$ \cite{Calabrese:2009qy}, which is insensitive to boundary condition $B$.

Performing the same computation for the expansion in the $B$-channel \eqref{eq:Bchann} gives 
\be  \label{eq:Bchannraw}
\bra{\psi_{AB} }\Phi_n (z, \bz)   \ket{\psi_{AB}}_{\text {UHP}} \approx C_0^B\alpha^{2h_n} (z \bz)^{- h_n } \[ 2 \sinh \frac{\alpha}{2} \(\log \frac{z}{\bz} - 2\pi i  \)\]^{-2h_n} \, .
 \ee
The coefficient $C_0^B$, up to a phase,   is the expectation value $B_{\Phi_n}$ of $\Phi_n$ in the UHP with  boundary conditions $B$  \cite{Lewellen:1991tb}. More precisely, mapping   to the strip  
\be  \label{eq:B3pt}
\bra{\psi_{AB} }\Phi_n (w, \bw)   \ket{\psi_{AB}}  \approx  
 B_{\Phi_n}  \( \frac{\alpha \pi }{2 L}\)^{2h_n}    \left[  \sin  \alpha \pi   \frac{L -\sigma}{ L}  \right]^{- 2h_n } \, .
\ee  
Here for $\sigma \to  L$ one recognises  the result for a small entangling region adjacent to the boundary at $\sigma=L$ with boundary conditions $B$  \cite{Calabrese:2009qy}. 

The correlator is then evaluated taking the largest between the approximate results  \eqref{eq:A3pt} and \eqref{eq:B3pt}, which yields
\be
 \label{eq:AB3pt}
\bra{\psi_{AB} }\Phi_n (w, \bw)  \ket{\psi_{AB}}  =  \( \frac{\alpha \pi }{2 L}\)^{2h_n} \max \left\{  A_{\Phi_n}   \left[  \sin  \alpha \pi   \frac{ \sigma }{ L}  \right]^{- 2h_n } ;  B_{\Phi_n}  \left[  \sin  \alpha \pi   \frac{L -\sigma}{ L}  \right]^{- 2h_n } \right\} \, .
 \ee
Substituting into  \eqref{eq:EE} and continuing the result to $n= 1$  gives the entanglement entropy. The maximization in equation~\eqref{eq:AB3pt} translates into a minimization over entanglement entropy contributions computed using the two channels
\be  \label{eq:EEstatic}
S = \frac{c}{6}  \log    \frac{2 L}{ \pi \alpha \epsilon } 
+\min \left\{ \frac{c}{6} \log  \sin \frac{   \alpha \pi  \sigma }{ L}  + s_A;  \frac{c}{6} \log  \sin  \frac{   \alpha \pi (L-\sigma) }{ L}  + s_B \right\}\, . 
\ee
In writing this result we have made explicit the UV regulator $\epsilon$ associated with the insertion of a replica twist operator $\Phi_n$. $s_{A}$ is the boundary entropy associated to the boundary with boundary conditions $A$ \cite{Affleck:1991tk,Calabrese:2004eu,Calabrese:2009qy}
\be
s_{A} \equiv \log g_{A} = \lim_{n\to1}\frac{1}{n-1} \log A_{\Phi_n} \, , 
\ee
and a similar definition holds for $s_B$.  
Here $g_{A,B}$ is the ground-state degeneracy  of  the corresponding boundary condition $A,B$ as defined in \cite{Affleck:1991tk}. Notice that the boundary entropy $s_{A,B}$ can be both positive or negative. In the holographic realisation of BCFTs it actually takes continuous values in the reals \cite{Takayanagi:2011zk,Fujita:2011fp} (see also the discussion in sec.~\ref{sec:holography}).

The result in  \eqref{eq:EEstatic}  reproduces  what found  in  \cite{Geng:2021iyq} using  a map to the twisted UHP where the effect of the stress energy tensor is trivialized.  
The case with homogeneous boundary condition $A$ and no BCC insertions \cite{Wen:2018vux} is recovered for $s_A=s_B$ and $\alpha=1$. In this limit  the two contributions coincide. Indeed, the correlator reduces to the twist operator one point function with boundary condition $A$, which is fixed by conformal invariance. 
Taking $s_A=s_B$ and  $0<\alpha <1$ in \eqref{eq:EEstatic}  can be regarded as  the  case of a  high energy eigenstates on the strip with homogeneous boundary condition $A$. This is  an excited state prepared by the insertion of a primary boundary operator $\psi$, rather than a BCC.   The conformal weight $h_\psi$ corresponds to the energy gap between this excited state and the ground state with homogeneous boundary conditions $A$.  For this state, the exchange of dominance between the two contributions as the size of the interval is changed  happens at $\sigma =L/2$.   

We are interested in the general case where $s_A \neq s_B$, and in holographic CFTs. For  these the boundary entropy scales like the central charge, $s_{A,B}\sim c$ and  $0<h_\psi/c \leq 1/24$  giving $0<\alpha \leq 1$ \cite{Biswas:2022xfw,Miyaji:2022dna}.   The  dominant channel is then determined by the boundary conditions $A$ and $B$  both via the boundary entropy $s_{A,B}$ and   the specific state,  through the energy gap \eqref{hBCC} encoded in $\alpha$.  
In particular, as the size $\sigma$ of the interval adjacent to $A$ is increased, $S$ undergoes a transition from channel $A$ to channel $B$ when  
\be  \label{eq:statictransition}
s_B - s_A  = \frac{c}{6} \log \frac{ \sin\(  \frac{\alpha \pi \sigma}{L} \) }{ \sin\(  \frac{\alpha\pi (L - \sigma)}{L}      \)}   \, . 
\ee
This happens for any  boundary state with  $s_A \neq s_B$ and for any value  $ 0< \alpha <1$. 
Notice  that the transition can happen for arbitrary small values of $\sigma$.  That is, channel $B$ can become dominant for $\sigma/ L\ll1$  as long as  $s_B-s_A$ takes a large negative value.  
Similarly, for  $s_B-s_A$ positive and sufficiently large one can remain in channel $A$ for $(L-\sigma)/L \ll1$.

\section{Entanglement entropy evolution under SSD and M\"obius Hamiltonians}\label{sec:EESSDM}

In the previous section we described the initial state of our system and the corresponding entanglement entropy computation. We now turn to the study of a particular class of quantum quenches. 

A quench is a sudden change in the system that produces a time-dependent excited state. Depending on the protocol, the change can affect the system locally or globally, homogeneously or inhomogeneously. Here we will focus on a class of inhomogeneous quenches that affect the system globally (for other studies on the topic see, \eg, \cite{Sotiriadis:2008ila,Calabrese:2016xau,Dubail:2016tsc,Alba:2021eni,Horvath:2021vlx,Balasubramanian:2013oga,Balasubramanian:2013rva,Sohrabi:2015qda,DeJonckheere:2018pbi,Moosavi:2019fas,Goto:2021sqx,Goto:2023wai,Nozaki:2023fkx,Liu:2023tiq}).
. They are obtained evolving the initial state with a  class of Hamiltonians known as sine-square-deformed (SSD) and M\"obius Hamiltonians \cite{Gendiar:2008udd,Hikihara:2011mtb,Katsura:2011ss,Gendiar:2010,Okunishi:2016zat,Wen:2016inm}. These are deformations of the usual Hamiltonian of the CFT defined on the strip, $H_0$. The state  $\ket{ \psi_{AB}}$ is an eigenstate of $H_0$ and represents an out-of-equilibrium state for the evolution dictated by these deformed Hamiltonians.

\subsection{SSD and M\"obius Hamiltonian on the strip} \label{ssec:Mobius}

The SSD Hamiltonian consists of an inhomogeneous spatial deformation of the standard Hamiltonian with a sine squared  enveloping function \cite{Wen:2018vux}
\be
H_{\rm SSD} \equiv \int_{0}^{L} d\sigma \,  2\sin^2\(\frac{\pi \sigma}{L} \) \, T_{\tau\tau}(\sigma)  \, . 
\ee
This can be expressed as 
\be
H_{\rm SSD} =   H_0 - \frac{1}{2} \( H_+ + H_{-}\)   
\ee
where 
\be \label{eq:H0strip}
H_0=  \int_{0}^{L} d\sigma  \,  T_{\tau\tau}(\sigma) =  \int_{0}^{L} \frac{dw}{2\pi} \,  \(T(w) +  \bT(\bw)\)
\ee
 is the standard CFT Hamiltonian on the strip, and the deformations  $H_{\pm}$  are defined as 
\be
 H_{\pm} =  \int_{0}^{L} \frac{dw}{2\pi} \(e^{\pm2\pi w / L}T(w) + e^{\mp 2\pi \bw / L}  \bT(\bw)\) \, .
\ee
The  M\"obius Hamiltonian generalizes this to a one parameter family of deformations
\be \label{eq:Mobius}
H_{\theta} \equiv  H_0 - \frac{\tanh (2\theta)}{2} \( H_+ + H_{-}\) \, , 
\ee
where the SSD Hamiltonian is recovered for $\theta \to \infty$.

Mapping the strip to the UHP by $z = e ^{\frac{\pi w}{L}}$  and using the properties of the stress tensor of a CFT in the UHP \cite{Cardy:2004hm}, one can write 
\be
\begin{aligned}
H_0&= 
 \frac{\pi}{L} \oint  z\, T(z) - \frac{\pi}{L} \frac{c}{24}   = \frac{\pi}{L}    \(L_0 -  \frac{c}{24}\)\,,  \\ 
 H_{+} + H_{-} &=   \frac{\pi}{L} \oint \(  z^{3} T(z)  + z^{-1} T(z) \)  = \frac{\pi}{L}  \(L_2+ L_{-2}\)\, ,
\end{aligned}
\ee
with the contour integral defined in the full complex plane (see appendix \ref{App:MIM} for details).

Next to the dilatation generator $L_0$, which defines with the Casimir  energy the standard Hamiltonian, $H_\theta$ involves the level-two Virasoro generators $L_{\pm 2}$. The subset  $\{ L_0, L_{\pm 2}\}$  forms a closed $sl(2,\mathbb{R})$ algebra  \cite{Witten:1987ty,Caputa:2022zsr}. It  is possible to write in an explicit closed form the action of the M\"obius Hamiltonian  $H_\theta$ on a primary operator $\mathcal O$  \cite{Okunishi:2016zat,Wen:2016inm,Wen:2018vux}.  Its Heisenberg evolution can be expressed as  a conformal  transformation to new coordinates $(z_\tE,\bz_\tE)$ such that 
\be \label{eq:HeisO}
\O_\tE(z,\bar z)  \equiv  e^{H_{\theta} \tE} \O(z,\bar z) e^{- H_{\theta} \tE}  =    \( \frac{\del z_\tE}{\del z} \)^{h_\O}  \( \frac{\del \bz_\tE}{\del \bz} \)^{\bar h_\O} \O (z_\tE,\bar z_\tE) \, .
\ee
The explicit form of the transformation can be obtained finding a map to a geometry where $H_{\theta}$ has a simple action    \cite{Okunishi:2016zat,Wen:2016inm,Wen:2018vux}. Using 
\be \label{eq:moebiusmap}
\tilde z^2 =   -\frac{ z^2 \cosh \theta  - \sinh \theta }{ z^2 \sinh \theta - \cosh \theta } \,
\ee
and a similar one for the $\bz$ coordinate, one gets  to
\be
\begin{aligned}
H_{\theta} &= \frac{ \pi}{L}\left[ \frac{1}{\cosh (2\theta) }   \oint  \tilde z T(\tilde z)  - \frac{c}{24}  \(1 -  \frac{ 1}{\cosh (2\theta) }  \) \right]  \\
&= \frac{ \pi}{L}\left[ \frac{1}{\cosh (2\theta) }  \tilde L_0   - \frac{c}{24}  \(1 -  \frac{ 1}{\cosh (2\theta) }  \) \right] \, .
\end{aligned}
\ee
Here $\tilde L_0$  the dilatation generator in the $\tilde z$-plane. The action of $H_{\theta}$ in the $\t z-$plane is then just a dilatation.
Assuming for simplicity ${h_\O}  ={\bar h_\O} $, which will be the relevant case for us here, one has 
\be
e^{H_{\theta} \tE} \O(\tilde z, \bar {\tilde  z}) e^{- H_{\theta} \tE}  =    \left| \frac{\del \tilde z_\tE}{\del \tilde z} \right|^{2 h_\O}   \O (\tilde z_\tE , \bar {\tilde  z}_\tE)  =  \lambda^{2 h_\O}  \O (\lambda \tilde z  , \lambda  \bar {\tilde  z})\, ,
\ee
where
\be
\t z_{t_E}  =  \lambda \t z \qquad\qquad \bar{ \t z}_{t_E}  =  \lambda \bar{\t z}\,  \label{eq:diltzt} \, , 
\ee
and with the scaling parameter $\lambda$ depending on the deformation $\theta$ and the Euclidean time $\tE$ 
\be \label{eq:lambdadef}
 \lambda  = \exp{ \frac{ \pi \, \tE}{L \cosh (2\theta) } }\, .
\ee
From here one can go back to the  $z$ coordinate using \eqref{eq:moebiusmap} on both sides of  \eqref{eq:diltzt}
\be
\( -\frac{ z^2  \cosh \theta- \sinh \theta }{ z^2 \sinh \theta - \cosh \theta } \)^{\frac{1}{2}} = \lambda \( -\frac{z_\tE^2 \cosh \theta  - \sinh \theta }{ z_\tE^2 \sinh \theta - \cosh \theta } \)^{\frac{1}{2}} \, .
\ee
The map for implementing the time dependence in the UHP is then more conveniently expressed as\footnote{As a consistency check, one can take the $t_E$-derivative on the two sides of equation \eqref{eq:HeisO},  and using the commutation relation between the Virasoro generators and the primary operator, one can match the coefficients mode by mode as in, \eg, \cite{Liska:2022vrd}.}
\be \label{eq:z2tau}
z_\tE^2 = \frac{ [(1-\lambda^2 ) \cosh (2 \theta )-(1+ \lambda^2 )] z^2 - (1-\lambda^2 ) \sinh (2 \theta)}{ (1-\lambda^2 )  \sinh (2 \theta )z^2  -(1-\lambda^2) \cosh (2 \theta )-(1+ \lambda^2)} \, .
\ee 
With a completely similar procedure it is immediate to write the result for the strip
\be \label{eq:wtau}
w_\tE = \frac{L}{2\pi}\log\left[ \frac{ [(1-\lambda^2 ) \cosh (2 \theta )-(1+ \lambda^2 )]   e^{\frac{2 \pi }{L}w } - (1-\lambda^2 ) \sinh (2 \theta)}{ (1-\lambda^2 )  \sinh (2 \theta ) e^{\frac{2 \pi }{L}w }    -(1-\lambda^2) \cosh (2 \theta )-(1+ \lambda^2)} \right] \, , 
\ee 
and analogous expressions  hold for  $\bz_\tE$ and $\bw_\tE $.

For later convenience, we write explicitly the form of the transformation in the SSD case, obtained as the $\theta \to \infty$ limit. The result is conveniently written in terms of $\tE$ as
\be \label{ztauSSD}
z_\tE^2 = \frac{ \frac{\pi \tE}{L}(z^2-1 ) + z^2}{\frac{\pi \tE}{L}(z^2-1 ) + 1} \, ,
\ee 
or equivalently for the strip  
\be \label{eq:wtauSSD}
w_\tE  = \frac{L}{2\pi} \log\left[ \frac{ \frac{\pi \tE}{L}(e^{\frac{2 \pi}{L}w}-1 ) + e^{\frac{2 \pi}{L} w}}{\frac{\pi \tE}{L}(e^{\frac{2 \pi}{L}w }-1 ) + 1} \right] \, .
\ee 
%

\subsection{Entanglement entropy after the  quench} \label{sec:Mobius} 

The Euclidean  post quench state is described by the density matrix
\be
\rho(\tE) = e^{- H_{\theta}\tE }  \ket{\psi_{AB}}\bra{\psi_{AB}}  e^{ H_{\theta}\tE} \, . 
\ee
The entanglement entropy for the interval $w \in [0, \sigma]$ in this state can be evaluated via the one-point function of a replica twist operator
\be \label{eq:Snevol}
S^{(n)}_\sigma = \frac{1}{1-n}\log \Tr \rho_\sigma^n =  \frac{1}{1-n}\log    \bra{\psi_{AB}}  e^{H_{\theta} \tE} \Phi_n(w=i \sigma,\bar w=- i\sigma) e^{-H_{\theta} \tE}    \ket{\psi_{AB}}  \, ,
\ee
analytically continuing the result to $n =1$ as in \eqref{eq:EE}.

The evaluation of the time evolved one-point twist correlator in \eqref{eq:Snevol} can be reduced to the computation in the UHP discussed in section \ref{sec:strip}. Mapping to the UHP and implementing the time evolution with  \eqref{eq:HeisO} one gets %
 \begin{align} 
  \bra{\psi_{AB}}  e^{H_{\theta} \tE} \Phi_n(w ,\bar w ) e^{-H_{\theta} \tE}    \ket{\psi_{AB}}  = &\left | \frac{\del z_\tE}{\del w} \right |^{2h_n}
      \langle  \psi_{AB} (\infty)    \Phi_n (z_\tE,\bar z_\tE)   \psi_{BA}(0) \rangle_{\rm UHP}\,   .  \label{eq:toevaluate} 
 \end{align}
This is analogus to \eqref{eq:3ptUHP}, with the crucial difference that here we are considering a time dependent state. This is reflected in the time dependent insertion point $ (z_\tE,\bar z_\tE) $ for the twist field and in the conformal factor, which is now  the composition of the map implementing the time dependence with the one relating the strip to the UHP. 
  
Following the approach used in the time independent case,   we will approximate $\langle  \psi_{AB} (\infty)    \Phi_n (z_\tE,\bar z_\tE)   \psi_{BA}(0) \rangle_{\rm UHP}$ \  with  the contribution coming from  the identity Virasoro block in the two channels  \eqref{eq:Achann}  and \eqref{eq:Bchann}. After continuing to real time  $\tE \to i t$,  the correlator will be evaluated as the dominant channel at each time $t$. The entanglement entropy, is then obtained as the analytic continuation of \eqref{eq:Snevol} to $n=1$.\\

In the next  subsection  we work out the explicit expression for the correlator \eqref{eq:toevaluate} and its time dependence. We do this first for the simpler SSD case and then for the M\"obius evolution. The reader interested in the physical results for the entanglement entropy can move directly to section~\ref{sec:EEevo}.

\subsubsection{Evaluation of the time dependent correlator}

\paragraph{SSD deformation.} We consider first the correlator \eqref{eq:toevaluate} approximated by the identity contribution to the $A$-channel expansion. Reading from \eqref{eq:AUHP} the UHP result in channel $A$ and analytically continuing to Lorentzian time with  $\tE \to i t $, we have
\begin{align}
 \bra{\psi_{AB}} e^{i H_{\text{SSD}} t} \Phi_n(w,\bar w) 
 &e^{-i H_{\text{SSD}} t}  \ket{\psi_{AB}} \no\\
 &\approx C_0^A \alpha^{2h_n} \( \frac{\del z_t}{\del  w} \)^{h_n} \( \frac{\del\bar z_t}{\del \bar w} \)^{h_n} (z_t \bz_t)^{(\alpha-1) h_n} ( z_t^\alpha - \bz_t^\alpha)^{-2h_n}  \, .
\end{align}
Making use of the explicit form of  $z_t$ and $\bz_t$ obtained from \eqref{ztauSSD}  with $z =e^{\frac{i \pi \sigma}{L}} $ and  $\tE \to i t$
\be
\begin{aligned} \label{ztSSD}
z_t = \frac{L \cos \frac{\pi \sigma} {L} - 2 \pi  t \sin \frac{\pi \sigma} {L} + i L \sin \frac{\pi \sigma} {L}}{\sqrt{(L \cos \frac{\pi \sigma} {L} -2 \pi  t \sin \frac{\pi \sigma} {L} )^2 +L^2 \sin^2 \frac{\pi \sigma} {L}}} \, , \\
\bar z_t= \frac{L \cos \frac{\pi \sigma} {L} + 2 \pi  t \sin \frac{\pi \sigma} {L} - i L \sin \frac{\pi \sigma} {L}}{\sqrt{(L \cos \frac{\pi \sigma} {L} +2 \pi  t \sin \frac{\pi \sigma} {L} )^2 - L^2 \sin^2 \frac{\pi \sigma} {L}}}  \, ,
\end{aligned}
\ee
gives
\be
\begin{aligned}
  \bra{\psi_{AB}} e^{i H_{\text{SSD}} t}  &\Phi_n(w,\bar w) e^{-i H_{\text{SSD}} t}  \ket{\psi_{AB}}    \\
  &=A_{\Phi_n}  \( \frac{ \pi \alpha }{2 L } \)^{2h_n}     \[  \(    f(t)^2 +   \sin ^2   \(  \frac{2 \pi  \sigma}{L}  \)   \)  \sin^2\(\frac{\alpha}{2} ~\delta(t) \)\]^{-h_n} \label{SSDt} \, ,
 \end{aligned}
 \ee
where we used the explicit relation  \eqref{eq:CA0} between  $C^A_0$  and $A_\Phi$.

In the above we have defined  the function 
\be \label{fSSD}
f(t) \equiv  - \frac{2 \pi ^2 t^2}{L^2}+\left(1+  \frac{2 \pi ^2 t ^2}{L^2}\right) \cos \left(\frac{2 \pi  \sigma  }{L}\right) \,,
\ee
and the phase $\delta(t)$ in \eqref{SSDt} is defined in terms of the ratio  
\be \label{phasedef}
e^{i \delta} \equiv \frac{z_t}{\bz_t} \, .  
\ee

\begin{figure}[t!]
\centering
  \includegraphics[width=0.33 \textwidth]{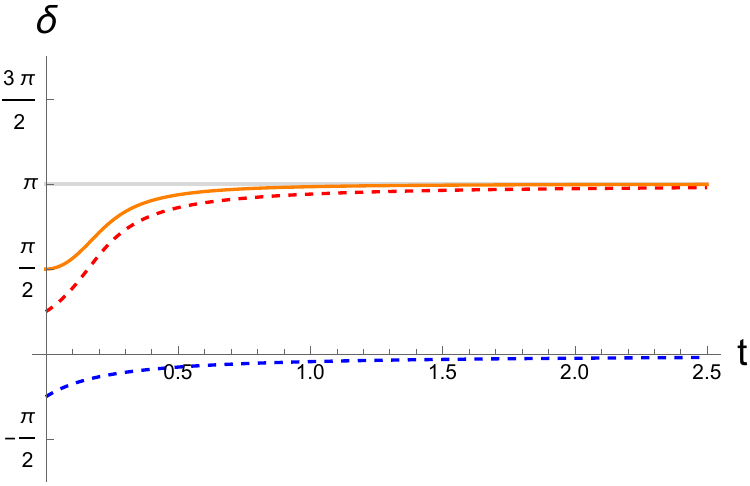} \hfill \includegraphics[width=0.33\textwidth]{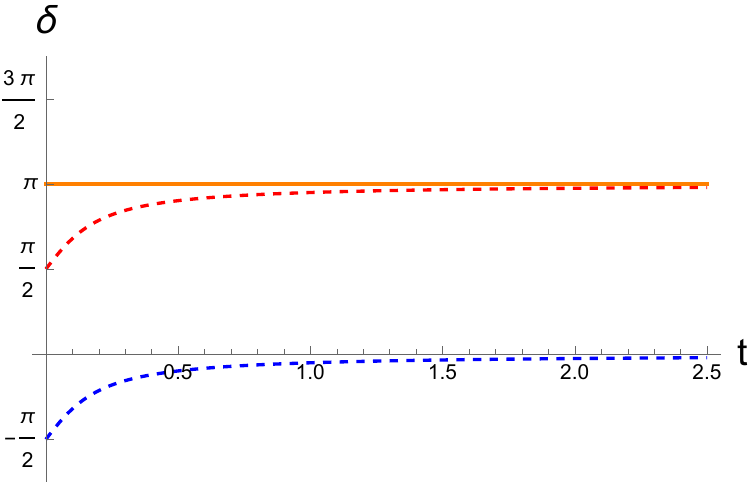}\hfill
  \includegraphics[width=0.33\textwidth]{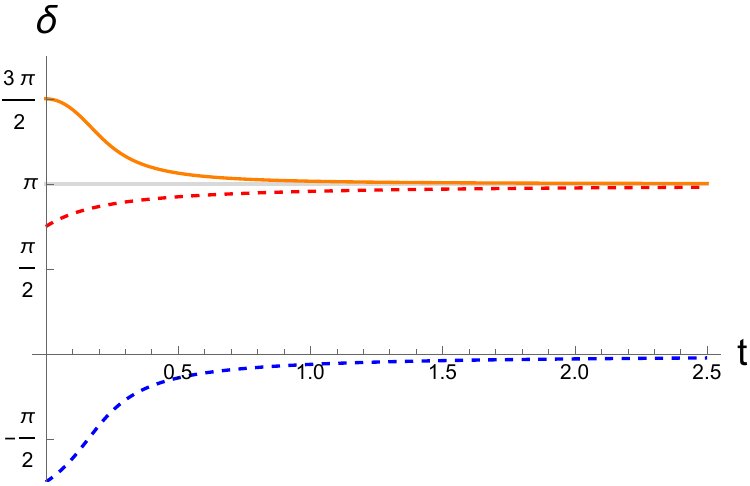}
\caption{SSD quench. Real time evolution of the phases of $z_t$ in  dashed red, $\bz_t$ in dashed blue, and $\delta(t)$ in solid orange. From left to right, $\sigma/L = \frac{1}{4}, \frac{1}{2}, \frac{3}{4}$.}
\label{fig:phasesSSD}
\end{figure}

The Lorentzian insertions $z_t$ and $\bar z_t$ in \eqref{ztSSD} move on the unit circle defining each a time dependent phase, which difference gives  $\delta(t)$.  In particular, $z_t$ goes from the initial value  $ z_{t=0} =e^{\frac{i \pi \sigma}{L}}$ anti-clock-wisely  to $e^{i \pi_+}$ for $t \to \infty$.  $\bar z_t$ goes  from   $\bz_{t=0}= e^{-\frac{i \pi \sigma}{L}}$  anti-clock-wisely to  $e^{i 0_{-}}$. The resulting ratio phase $\delta(t)$ goes from $2 \pi \sigma /L$ to $\pi/L$ monotonically from below (above) for $\sigma < L/2$   ($\sigma> L/2$  ). This is illustrated in   fig.~\ref{fig:phasesSSD}. The result can also be expressed in terms of the function $f(t)$ defined in \eqref{fSSD} as 
\be  \label{eq:dSSD}
e^{i   \delta} = \frac{  f(t ) +  i  \sin    \( \frac{2 \pi  \sigma}{L}   \) }{  \sqrt{f(t ) ^2 +   \sin ^2   \( \frac{2  \pi  \sigma}{L}   \)}}
 \, .
\ee
Notice that in the limiting case $\sigma =L/2$ there is actually  no time evolution for the phase $\delta(t)$.
Despite  $z_t$ and   $\bz_t$ changing in time, their relative  distance along the unit circle remains constant and equal to $\pi$.  The  factor  $f(t)^2 +   \sin ^2   \(  \frac{2 \pi  \sigma}{L}  \) $ in \eqref{SSDt} has nevertheless a non-trivial evolution, which yields the time dependence for the full correlator.

As discussed in section~\ref{sec:strip}, one can perform a different expansion for the three-point  function and approximate the result with the $B$-channel Virasoro identity block \eqref{eq:Bchannraw}. 
The $A$-channel and   $B$-channel blocks differ by a monodromy, which however does not affect the conformal factor in \eqref{eq:toevaluate}. The  results for the $B$-channel follows in a straightforward manner from \eqref{eq:Bchannraw} and is obtained  performing the replacements $\delta(t) \to 2\pi -  \delta(t)$ and  $A_{\Phi_n}\to B_{\Phi_n}$  in \eqref{SSDt}
\begin{align}
\label{SSDtB}
 \bra{\psi_{AB}} &e^{i H_{\text{SSD}} t} \Phi(w,\bar w) e^{-i H_{\text{SSD}} t}  \ket{\psi_{AB}}    \no\\
&  =  B_{\Phi_n}   \( \frac{ \pi \alpha }{2 L } \)^{2h_n}    \[  \(    f(t)^2 +   \sin ^2   \(  \frac{2 \pi  \sigma}{L}  \)   \) \sin^2\(\frac{\alpha}{2} \( 2 \pi  - \delta(t)\) \) \]^{-h_n}\, . 
\end{align}

\paragraph{M\"obius deformation.} For the M\"obius case, the evaluation of the correlator  \eqref{eq:toevaluate}  is performed with a  similar computation. 

The relevant coordinate transformation implementing the Euclidean evolution is now  \eqref{eq:z2tau}. Continuing to Lorentzian time  $\tE \to i t$ and  defining for compactness the rescaled time variable 
\be \label{eq:Teff}
 T \equiv   \frac{ \pi t }{L \cosh (2 \theta)  }\, , 
\ee
one obtains
\be \label{eq:ztimeM}
z_t= \frac{\cos (T) \cos \left(\frac{\pi  \sigma }{L}\right)-e^{2 \theta } \sin (T) \sin \left(\frac{\pi  \sigma }{L}\right)+i \left(e^{-2 \theta } \cos \left(\frac{\pi  \sigma }{L}\right) \sin (T)+\cos (T) \sin \left(\frac{\pi  \sigma  }{L}\right)\right)}{\sqrt{\cos ^2(T)-\sin (2 T) \sin \left(\frac{2 \pi  \sigma }{L}\right) \sinh (2 \theta )+\sin ^2(T)   \left(\cosh (4 \theta )-\cos \left(\frac{2 \pi  \sigma }{L}\right) \sinh (4 \theta )\right)}}\,. 
\ee
with the expression for $\bz_t$   obtained via the replacement $\sigma \to - \sigma$. 

The correlator approximated by the identity block contribution in the $A$-channel can be expressed in the form  
\begin{align}
\bra{\psi_{AB} } &e^{i H_{\theta} t} \Phi(w,\bar w) e^{-i  H_{\theta} t}  \ket{\psi_{AB} } \no\\
&=    A_{\Phi_n}  \( \frac{ \pi \alpha }{2 L } \)^{2h_n}    \left[   \(     f_\theta(T)^2 +   \sin ^2   \( \frac{2 \pi  \sigma}{L}   \)   \)   \sin^2 \( \frac{\alpha}{2} \delta_{\theta}(T) \)   \right]^{-h_n} \,  , 
 \end{align}
while the in $B$-channel it reads
\begin{align}
 \bra{\psi_{AB} } & e^{i H_{\theta} t} \Phi(w,\bar w) e^{-i  H_{\theta} t}  \ket{\psi_{AB} } \no\\
 &=    B_{\Phi_n} \( \frac{ \pi \alpha }{2 L } \)^{2h_n}    \left[   \(     f_\theta(T)^2 +   \sin ^2   \( \frac{2 \pi  \sigma}{L}   \)   \)   \sin^2 \( \frac{\alpha}{2} \(2\pi -\delta_{\theta}(T)\)  \)   \right]^{-h_n} \, . 
\end{align}

 The difference with respect to the SSD case  completely resides in the explicit form of the function $f_\theta$ and phase $\delta_\theta$. In particular, 
\be \label{eq:ftheta}
f_\theta(T) \equiv -  \sin^2(T) \sinh (4 \theta )+\left(\cos ^2(T)+ \cosh (4 \theta ) \sin^2(T)\right) \cos \left(\frac{2 \pi  \sigma }{L}\right)
\ee
is the finite $\theta$ generalization of the function $f(t)$ defined above for the SSD case and recovered as the $\theta \to \infty$ limit. The phase $\delta_\theta(T)$ is similarly defined as
\be  
e^{i \delta_\theta} \equiv  \frac{z_t}{\bz_t} \,   
 \ee
with $z_t$  and  $\bz_t$ given in \eqref{eq:ztimeM}.

\begin{figure}[t!]
\centering
  \includegraphics[width=0.33 \textwidth]{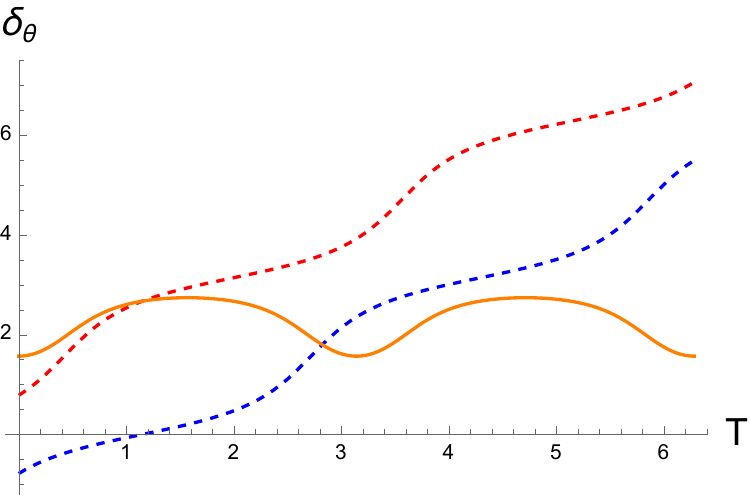} \hfill \includegraphics[width=0.33\textwidth]{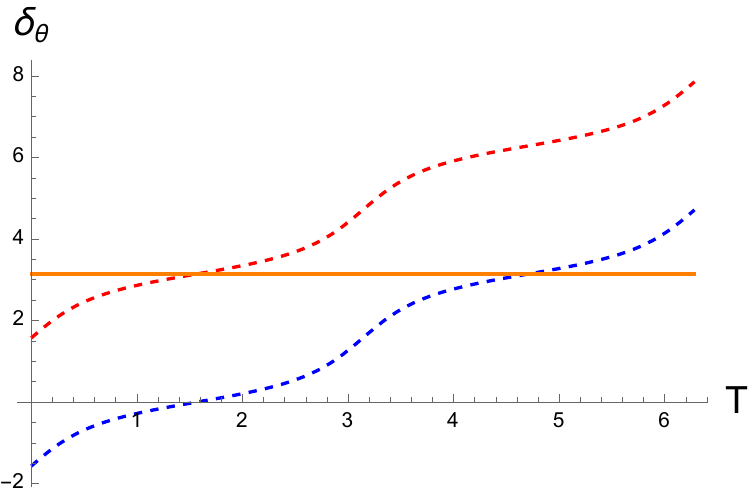}\hfill
  \includegraphics[width=0.33\textwidth]{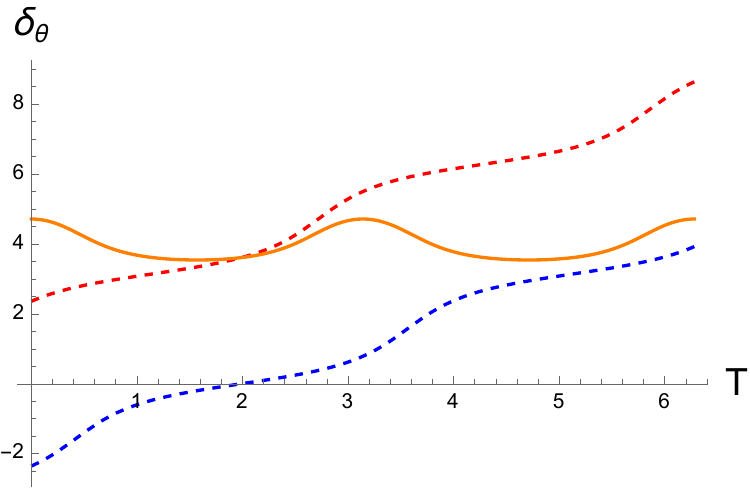}
\caption{M\"obius quench. Real time evolution of the phases of $z_t$ in dashed red, $\bz_t$ in dashed blue, and $\delta(t)$ in solid orange. From left to right, $\sigma/L = \frac{1}{4}, \frac{1}{2}, \frac{3}{4}$. In all plots $\theta=0.4$.  }
\label{fig:phasesM}
\end{figure}

As time evolves, $z_t$  and  $\bz_t$  move around the unit circle with $T$ periodicity $2\pi$. Within half a period, $z_t$ goes  from the initial value $ z_{t=0} =e^{\frac{i \sigma \pi}{L}}$ to  $e^{ i \pi  + \frac{i \sigma \pi}{L}}$, and  $\bz_t$ from $ \bz_{t=0} =e^{-\frac{i \sigma \pi}{L}}$ to  $e^{ i \pi  - \frac{i \sigma \pi}{L}}$, as portrayed in fig.~\ref{fig:phasesM}. 
Despite the individual phases of  $z_t$ and  $\bz_t$  always increase in time, their difference yields an oscillatory behaviour for $\delta_\theta(T)$  with periodicity $ \pi$, as shown explicitly in fig.~\ref{fig:phasesM}. The time evolution of    $\delta_\theta(T)$  can be expressed as  
\be  \label{eq:deltatheta}
e^{i   \delta_{\theta}} = \frac{  f_\theta(T ) +  i  \sin    \( \frac{2 \pi  \sigma}{L}   \)  }{ \sqrt{ f_\theta(T ) ^2 +   \sin ^2   \( \frac{2  \pi  \sigma}{L}   \) }}\, . 
\ee
 $\delta_\theta$   starts from the initial value $ \frac{2  \pi \sigma}{L}$  and reaches an extremal  value 
\be \label{dmid}
\delta_{\text{ext}}  = \frac{2 \pi \sigma}{L} +2 \arctan\left( \frac{\sin \( \frac{2\pi \sigma}{L} \)}{\coth (2 \theta )-\cos \( \frac{ 2\pi \sigma}{L} \)  } \right)  \,
\ee
for  $T=\pi/2$, before going back to the initial value. $\delta_{\text{ext}}$ represents a  maximum for $\sigma < L/2$ and a minimum  for  $\sigma > L/2$, as visible in fig.~\ref{fig:phasesM}. 
\begin{figure}[t!]
\centering
  \includegraphics[width=0.4 \textwidth]{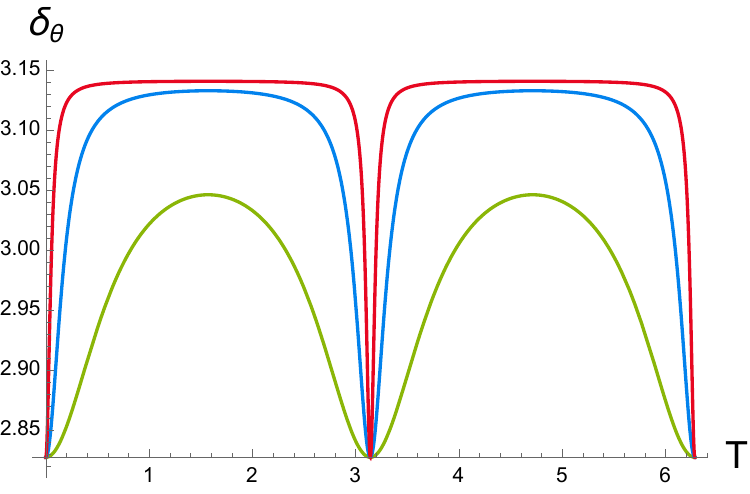} \qquad\qquad \includegraphics[width=0.4\textwidth]{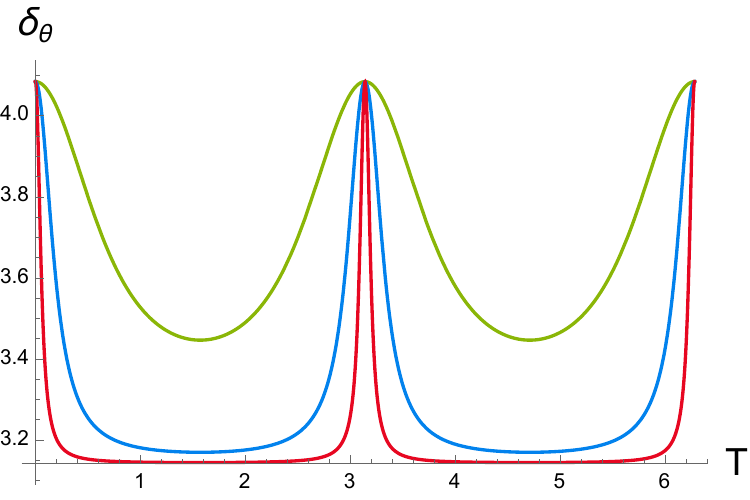}\\ 
  \includegraphics[width=0.4 \textwidth]{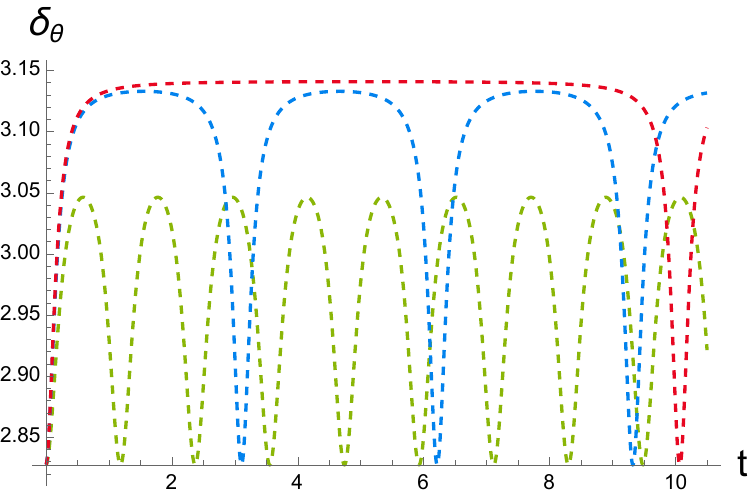} \qquad \qquad\includegraphics[width=0.4\textwidth]{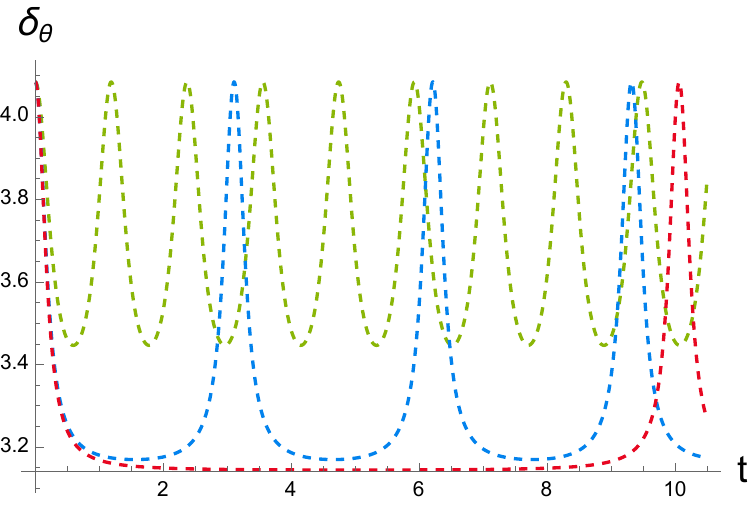}
 \caption{M\"obius quench. The  oscillatory behaviour for $\delta_{\theta}$ is depicted as a function of the effective time $T$  and physical time   $t$, for $\sigma/L =0.45 $ (left) and  $\sigma/L =0.65 $ (right). $\theta$ = 0.3 (green), 0.9 (blue), 1.5 (red).
}
\label{fig:thetadepM}
\end{figure}

For fixed $\sigma$, increasing values of $\theta$ give values of $\delta_{\text{ext}}$ closer to $\pi$ (see fig.~\ref{fig:thetadepM}). 
In terms of the physical time  $t$, the oscillatory period increases with $\theta$ yielding an effective description where the physical size of the system is rescaled to $L_{\rm eff} \sim L \cosh (2\theta)$ (see, \eg, \cite{Wen:2018vux}). In the $\theta\to \infty$ limit the effective size goes to infinity, suppressing the return phase, and $\delta_{\text{ext}} \to \pi$  giving back the SSD evolution in fig.~\ref{fig:phasesSSD}.  %
 
\subsubsection{Evolution of the entanglement entropy} \label{sec:EEevo}
 
We summarize the result obtained in the previous subsection for the twist one-point function  \eqref{eq:toevaluate} in the time dependent state and discuss the ensuing entanglement entropy dynamics. We start considering the SSD case and then move to the M\"obius quench.

\paragraph{SSD quench.} Under the working assumption that the correlator \eqref{eq:toevaluate}  is well approximated by the identity Virasoro block in the $A$- or $B$-channel expansion, in Lorentzian time we have  
\begin{align}
 \label{eq:AB3ptquench}
\bra{\psi_{AB} }\Phi_n (w, \bw)  \ket{\psi_{AB}}  =  &\( \frac{\alpha \pi }{2 L}\)^{2h_n}   \(     f(t)^2 +   \sin ^2   \( \frac{2 \pi  \sigma}{L}   \)   \)^{-h_n}  \\
&\times \max \left\{   A_{\Phi_n}       \left[     \sin^2 \( \frac{\alpha}{2} \delta (t) \)   \right]^{-h_n}  ; B_{\Phi_n}       \left[     \sin^2 \( \frac{\alpha}{2}( 2 \pi - \delta(t) )  \)   \right]^{-h_n}   \right\} \, ,  \nonumber
\end{align} 
with $f(t)$  given in \eqref{fSSD} and $\delta(t)$  in \eqref{eq:dSSD}. 
  
Plugging \eqref{eq:AB3ptquench} into the expression for the R\'enyi entropies \eqref{eq:Snevol} and continuing to $n=1$ gives the entanglement entropy for $t>0$ 
\begin{align} \label{eq:CFTresult}
S(t)= \frac{c}{12}  \log &  \[ \(  \frac{2 L}{ \pi \alpha \epsilon } \)^2 \(     f^2(t)+   \sin ^2   \( \frac{2 \pi  \sigma}{L}  \)  \)  \]\\
&+\min \left\{
 \frac{c}{12}  \log  \sin^2\(\frac{\alpha}{2}   \delta(t)  \)  + s_A;  \frac{c}{12}  \log   \sin^2\(\frac{\alpha}{2} \(2 \pi   -   \delta(t)  \)  \) + s_B\right\} \, ,  \nonumber
\end{align}
where we have reinstated the UV regulator $\epsilon$. 

$S(t)$ is thus obtained at each time minimizing over the two competing terms inside the brackets, corresponding to the two channels.   Depending on the values of the different parameters, one channel can dominate for all times or there can be a transition from one to the other contribution at some finite time $t^*$.  

Consider first a situation where $s_A = s_B$ and   $0<\alpha < 1$. This is interpreted as the case where the initial state is a high energy eigenstate of the undeformed Hamiltonian on the strip with equal boundary conditions (see section \ref{sec:univSEE}). In this case, the  initial value $ \delta(t=0) =  2 \pi \sigma /L$ determines which of the two contributions in \eqref{eq:CFTresult} dominates for all times. In fact, the condition to have an exchange of dominance and a transition from the initial channel to the other is to cross the value $ \delta(t) = \pi$ at some finite time $t^*$. However this is only reached asymptotically for  $t\to \infty$ in the SSD case (see fig.~\ref{fig:phasesSSD}).
The resulting evolution is then qualitatively similar to the case of the  ground state with equal boundary conditions on the two sides of the strip, studied in \cite{Wen:2018vux}.\footnote{This is recovered for $\alpha =1$, in which case the two contributions in \eqref{eq:CFTresult} coincide.} Indeed the late time growth is $S(t) \sim \frac{c}{3}\log t$ and does not exhibit any revival, compatibly with an effective  infinite length for the deformed dynamics. It also admits a quasi-particle interpretation, as detailed in \cite{Wen:2018vux}.

\begin{figure}[t!]
\centering
\includegraphics[width=0.4\textwidth]{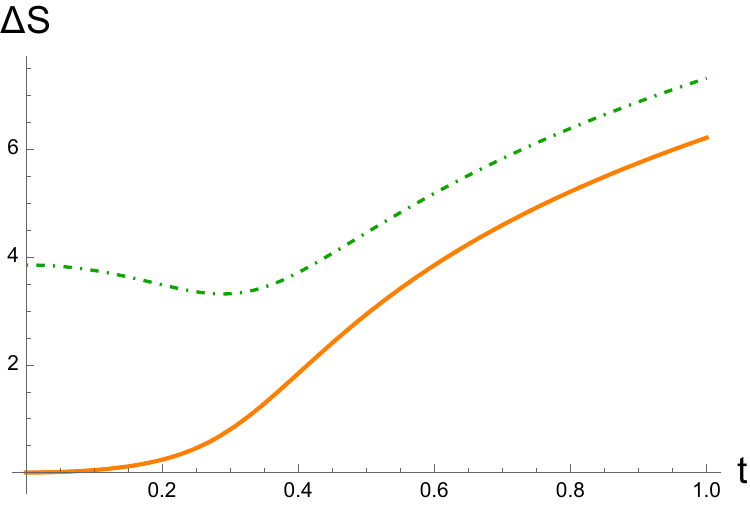} \qquad \qquad \includegraphics[width=0.4\textwidth]{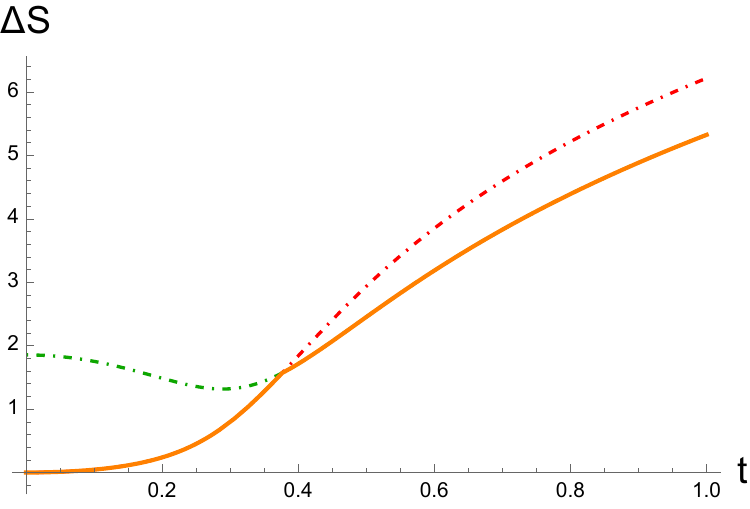} \\
\includegraphics[width=0.4 \textwidth]{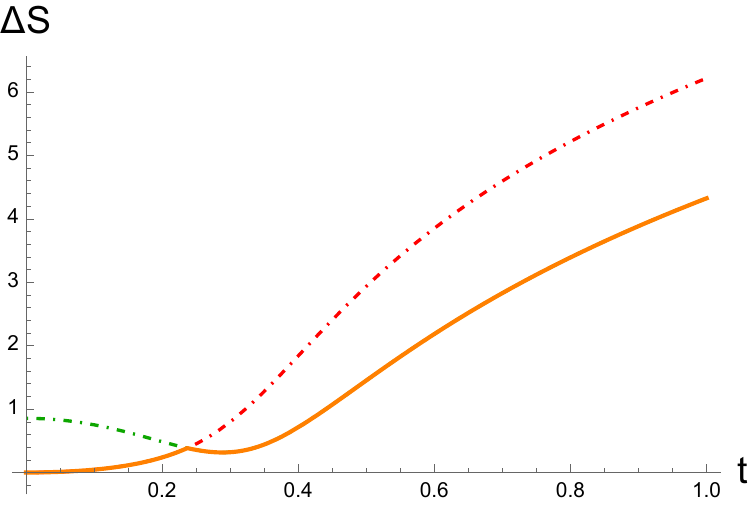} \qquad \qquad 
\includegraphics[width=0.4\textwidth]{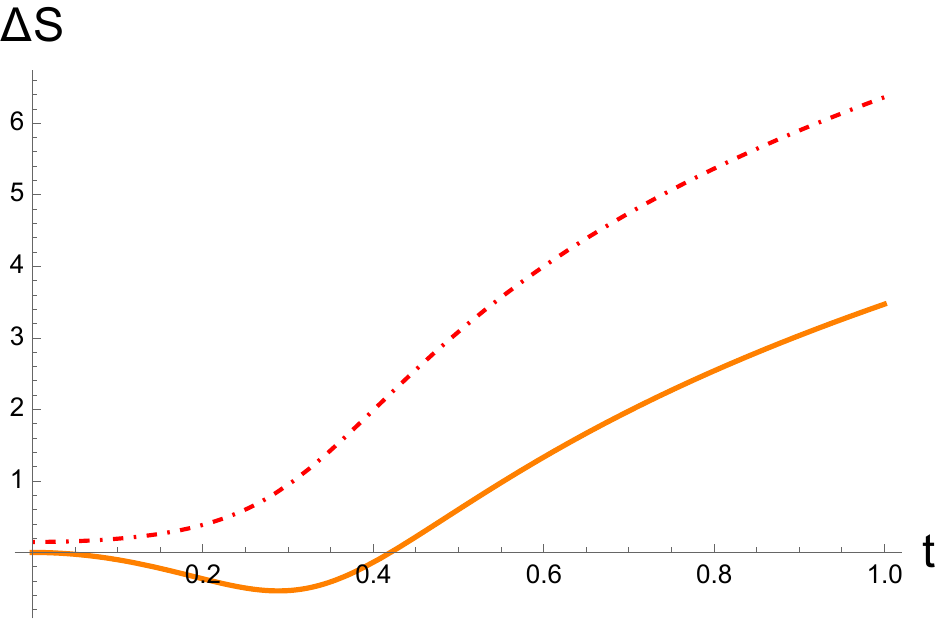} 
\caption{Plot of the entanglement entropy \eqref{eq:CFTresult}, rescaled by $12/c$, with the initial value subtracted. The solid line is  the entanglement entropy obtained at each time with the minimization procedure over the two competing channels in \eqref{eq:CFTresult}. The dot-dashed lines correspond to the two competing channels in their non-dominant phase, red for channel $A$ and green for channel $B$. 
The plots show the case of $\sigma/L = 0.15$ and $\alpha=0.5$, for $ \frac{12}{c}(s_B-s_A) = 1,-1,-2,-3$ from top-left to bottom-right. 
 }
\label{fig:transitionSSD}
\end{figure}

For general values of  $s_A$ and  $s_B$, the late time behaviour is unchanged,  but at intermediate times the situation is  richer and a dynamical exchange of dominance between channel $A$ and $B$ may occur. 
Figure~\ref{fig:transitionSSD} illustrates an example of how different values of the boundary entropy give  different patterns.

In order to have a transition the two contributions in \eqref{eq:CFTresult} should exchange dominance at some time  $t^*$. In other words, 
\be \label{eq:transition}
\frac{c}{6} \log  \frac{ \sin \(\frac{\alpha}{2} \delta(t)   \)}{ \sin \(\frac{\alpha}{2} \( 2\pi - \delta(t) \)    \)}  = s_B-  s_A \, . 
\ee
must have a solution at some value of time $t^*$. In the SSD case this condition is not satisfied for arbitrary values of  $s_A$ and  $s_B$. 

The left hand side of \eqref{eq:transition} is always negative and a monotonically increasing function of $\delta$. 
 For $0 <\sigma \leq L/2$, $\delta(t)$ increases monotonically from $2\pi \sigma/L$ at $t=0$ to $\pi$ for $t \to \infty$. To observe a transition at some finite time $t^*$,  the boundary entropies must then fall within the range
\be\label{eq:rangeSSD1}
\frac{c}{6} \log \frac{ \sin\(  \frac{\alpha \pi \sigma}{L} \) }{ \sin\(  \frac{\alpha\pi (L - \sigma)}{L}      \)} < s_B - s_A   < 0 \ .
\ee
The monotonicity of $\delta(t)$ also implies that when this condition is satisfied, the initial value of the entropy corresponds to the contribution given by channel $A$. Therefore the transition will be from channel $A$ to channel $B$ as time  increases. On the other hand, when the inequality is violated there is no transition and the dominant channels are 
\be
\begin{aligned}
A\text{-channel}&~~\text{for} ~~ s_B - s_A>0 \,\\
B\text{-channel}&~~\text{for} ~~ s_B - s_A  < \frac{c}{6} \log \frac{ \sin\(  \frac{\alpha \pi \sigma}{L} \) }{ \sin\(  \frac{\alpha\pi (L - \sigma)}{L} \)}   . 
 \end{aligned}
\ee
 Notice that in the limiting case $\sigma = L/2$ the window of values \eqref{eq:rangeSSD1} for  the boundary entropies  closes and there is no dynamical transition despite having a time evolving entanglement entropy.

For  $L/2<\sigma<L$  the situation is similar, but $\delta(t)$ now decreases monotonically in time from $2\pi \sigma/L$ to $\pi$.
In practical terms, the dynamics can be described inverting $A$ and $B$ and a transition from channel B to channel A happens for 
\be
0 < s_B - s_A   <\frac{c}{6} \log \frac{ \sin\(  \frac{\alpha \pi \sigma}{L} \) }{ \sin\(  \frac{\alpha\pi (L - \sigma)}{L}      \)}  \,.
\ee
The complementary range of  $s_B - s_A$, where there is no transition, can instead be summarized as
\be
\begin{aligned}
B\text{-channel}&~~\text{for} ~~ s_B - s_A< 0 \,\\
A\text{-channel}&~~\text{for} ~~ s_B - s_A  > \frac{c}{6} \log \frac{ \sin\(  \frac{\alpha \pi \sigma}{L} \) }{ \sin\(  \frac{\alpha\pi (L - \sigma)}{L} \)}   . 
 \end{aligned}
\ee

As shown in fig.~\ref{fig:transitionSSD},  for $0 <\sigma \leq L/2$,  the transition happens at earlier times for smaller values of  $s_B-s_A$. In fact, $\delta(t)$ is a monotonically increasing function of time and solving \eqref{eq:transition} for $\delta(t^*)$ gives the relation
\be \label{deltatransition}
\cot \( \frac{\alpha}{2}  \delta(t^*)\) = \cot(\pi \alpha) + \frac{e^{ 6(s_B-s_A)/c}}{\sin(\pi \alpha)}  \, ,
\ee
On the contrary, for  $L/2<\sigma<L$ the transition happens at larger times for decreasing values of $s_B-s_A$.

\paragraph{M\"obius deformation.}  The M\"obius  result for the time dependent three point function  and the corresponding  entanglement entropy  can be written in the same form as for the SSD quench
\begin{align}
 \label{eq:AB3ptquenchM}
\bra{\psi_{AB} }\Phi_n (w, \bw)  \ket{\psi_{AB}}  =  &\( \frac{\alpha \pi }{2 L}\)^{2h_n}   \(     f_\theta(T)^2 +   \sin ^2   \( \frac{2 \pi  \sigma}{L}   \)   \)^{-h_n}   \\
&\times \max \left\{   A_{\Phi_n}       \left[     \sin^2 \( \frac{\alpha}{2} \delta_\theta (T) \)   \right]^{-h_n}  ; B_{\Phi_n}       \left[     \sin^2 \( \frac{\alpha}{2}( 2 \pi - \delta_\theta (T))  \)   \right]^{-h_n}   \right\} \, ,\nonumber
\end{align} 
and 
\begin{align} \label{eq:CFTresultM}
S(T)= \frac{c}{12}  \log &  \[ \(  \frac{2 L}{ \pi \alpha \epsilon } \)^2 \(     f_\theta(T)^2+   \sin ^2   \( \frac{2 \pi  \sigma}{L}  \)  \)  \]\\
&+\min \left\{
 \frac{c}{12}  \log  \sin^2\(\frac{\alpha}{2}    \delta_\theta (T) \)  + s_A;  \frac{c}{12}  \log   \sin^2\(\frac{\alpha}{2} \(2 \pi   -    \delta_\theta (T)  \)  \) + s_B\right\} \, . \nonumber
\end{align}
The difference resides in the definition of the function $f_\theta$, in \eqref{eq:ftheta}, and in the time oscillating  phase $\delta_\theta(T)$ (see equation \eqref{eq:deltatheta} and discussion around it).

Let us illustrate the case  $0 <\sigma \leq L/2$  for simplicity. The physics in the complementary range of $\sigma$ is easily obtained and substantially amounts to interchanging $A$ and $B$ in the following discussion. 

 The phase $\delta_\theta(T)$  oscillates with periodicity $\pi$ between the initial value  $2\pi \sigma/L$ and the extremal, maximum, value $\delta_{\text {ext}}<\pi$ in \eqref{dmid} (see fig.~\ref{fig:thetadepM}).  To have a dynamical phase transition between the two channels in  \eqref{eq:CFTresultM} the boundary entropies difference must fall within the range
\be\label{eq:rangeM1}
\frac{c}{6} \log \frac{ \sin\(  \frac{\alpha \pi \sigma}{L} \) }{ \sin\(  \frac{\alpha\pi (L - \sigma)}{L}      \)} < s_B - s_A  < \frac{c}{6} \log  \frac{ \sin\(\frac{\alpha}{2} \delta_{ \text {ext}}   \)}{ \sin\(\frac{\alpha}{2} \( 2\pi -  \delta_{ \text {ext}}  \)    \)}  <0 \ .
\ee
As compared to the SSD case, the  range of $s_B - s_A$  which leads to a transition is reduced. Smaller values of the deformation parameter $\theta$ correspond to a smaller range of $s_B - s_A$ available for the transition to happen.  In the limit  $\theta\to0$,  the transition window in \eqref{eq:rangeM1} closes reflecting the fact that there is no dynamics when the deformation is off. 
 
\begin{figure}[t!]
\centering
  \includegraphics[width=0.32 \textwidth]{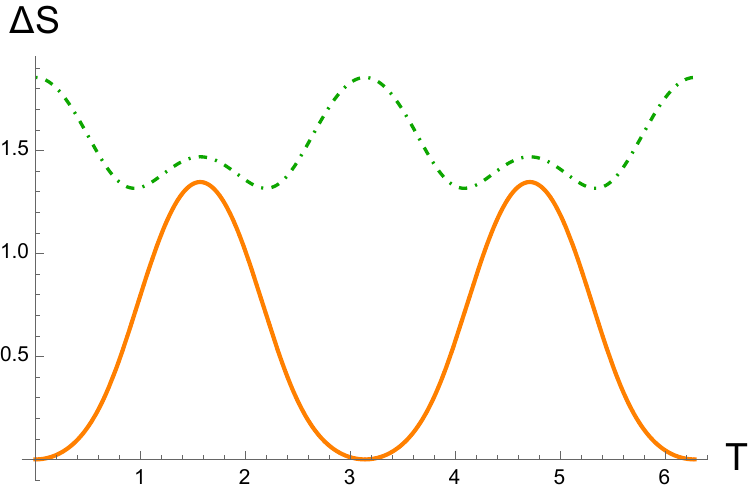} \hfill \includegraphics[width=0.32\textwidth]{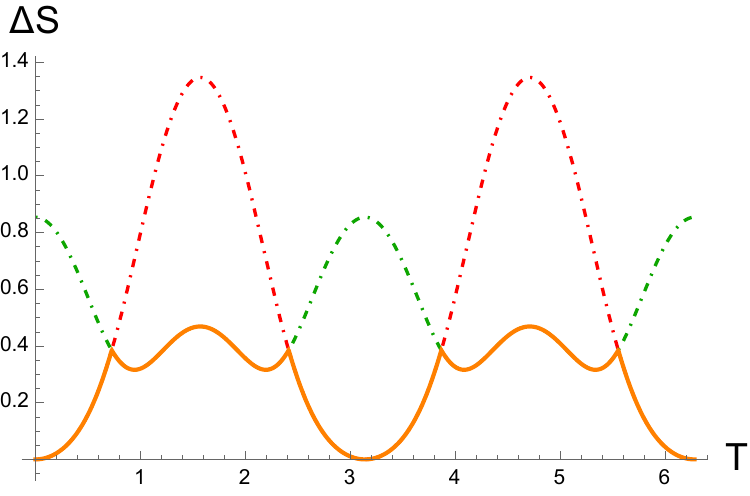} \hfill \includegraphics[width=0.32\textwidth]{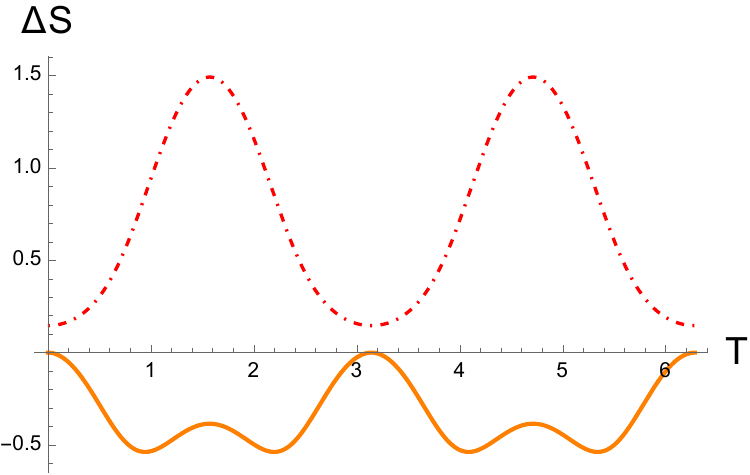}\\  
 \caption{Plot of the entanglement entropy, rescaled by $12/c$,  with the initial value subtracted. In all plots $\sigma/L= 0.15, \alpha =0.5$ and $\theta =0.3$. From left to right $ \frac{12}{c}(s_B-s_A) =-1,-2,-3$. The dot-dashed lines correspond to channel A (red) and B (green) when in their non-dominant phase. The solid line is the entanglement entropy.  The dominant channel is $A$ in the left plot and $B$ in the right plot. In the middle there is a transition from channel $A$ to $B$ and back to $A$ happening within each period.}
\label{fig:EEMsdep}
\end{figure}

An example of the  entanglement entropy evolution  following  from the minimization procedure  is depicted   in fig.~\ref{fig:EEMsdep}. The different panels show how depending on the value of $s_B - s_A$  there is or not a transition.

Similarly to the SSD case, outside the range \eqref{eq:rangeM1} there is no transition and the term within brackets in \eqref{eq:CFTresultM}  is minimized by the
\be
\begin{aligned}
A\text{-channel}&~~\text{for} ~~ s_B - s_A>\frac{c}{6} \log  \frac{ \sin\(\frac{\alpha}{2} \delta_{ \text {ext}}   \)}{ \sin\(\frac{\alpha}{2} \( 2\pi -  \delta_{ \text {ext}}  \)    \)} \, ,\\
B\text{-channel}&~~\text{for} ~~ s_B - s_A  < \frac{c}{6} \log \frac{ \sin\(  \frac{\alpha \pi \sigma}{L} \) }{ \sin\(  \frac{\alpha\pi (L - \sigma)}{L} \)}   . 
 \end{aligned}
\ee

\begin{figure}[t!]
\centering  \includegraphics[width=0.45 \textwidth]{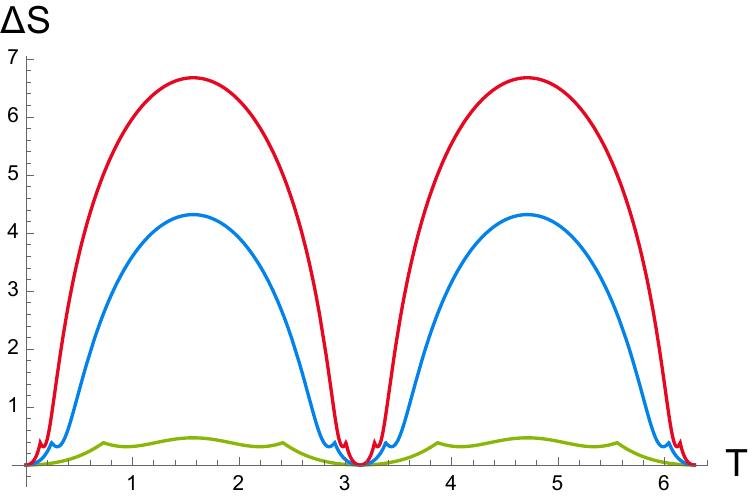} \hfill \includegraphics[width=0.45\textwidth]{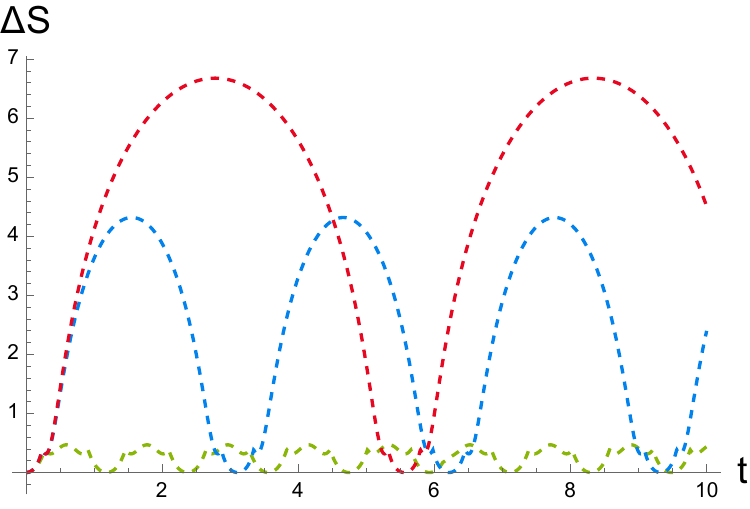}
 \caption{Plot of the entanglement entropy with the initial value subtracted and rescaled by $12/c$.  The plots are for $\sigma/L= 0.15,\alpha =0.5$ and $s_B-s_A = - \frac{c}{6}$. $\theta$ = 0.3 (green), 0.9 (blue), 1.5 (red).  The plot in terms of the physical time $t$ displays an evolution compatible with an effective spatial size $L_{\text{eff}}\sim L \cosh ( 2\theta) $.}
\label{fig:EEM}
\end{figure}

In fig.~\ref{fig:EEM}  we also report sample plots of the time  dependence of the entanglement entropy  for different values of $\theta$.  

One important difference with respect to the SSD quench is the oscillatory character of the dynamics that characterizes  the M\"obius case. When present, the  transition happens at times $T^*$  where 
\be 
\frac{c}{6} \log  \frac{ \sin\(\frac{\alpha}{2} \delta_\theta(T^*)   \)}{ \sin\(\frac{\alpha}{2} \( 2\pi - \delta_\theta(T^*)\)    \)}  = s_B-s_A \, .
\ee
As seen explicitly  in the central panel of fig.~\ref{fig:EEMsdep}, when a first transition from channel $A$ to channel $B$ happens at a certain time $T^*$ this will be necessarily followed by a transition back to channel $A$ at a time $\pi - T^*$, and so forth for each period of the effective time $T$.  \\

\section{Quench with generally deformed Hamiltonians}\label{sec:GeneralDeform}

 We have discussed quenches induced by the SSD  and  M\"obius Hamiltonians. On the strip these are constructed through a deformation of the usual CFT Hamiltonian given in terms of  level-two Virasoro generators $L_{\pm 2}$.  
 
 This construction can be extended to define a general class of deformed Hamiltonians obtained from any subset of the Virasoro generators  $\{ L_0, L_{\pm k}\}$  with $k\in \mathbb{Z}^+$ (see, \eg, \cite{Wen:2018agb,Wen:2020wee,Fan:2020orx,Wen:2021mlv}), which also form a $sl(2,\mathbb{R})$ algebra \cite{Witten:1987ty,Caputa:2022zsr}.  In this section we extend our analysis  to this general class of deformed Hamiltonians.

\subsection{A general class of deformed Hamiltonians}

 In analogy with  the M\"obius Hamiltonian \eqref{eq:Mobius}, the general class of  deformed Hamiltonians constructed from $\{ L_0, L_{\pm k}\}$  can be written as
\be\label{eq:generalH}
H_k \equiv H_0 - \frac{\tanh(2\theta)}{2} (H^{+}_k + H^-_k) 
\ee
where $H_0$ is the undeformed Hamiltonian \eqref{eq:H0strip} on an interval of width $L$,  and the deformations $H^{\pm}_k$ are defined as
\be 
 H^{\pm}_{k} =  \int_{0}^{L} \frac{dw}{2\pi} \(e^{\pm k \pi w / L}  T(w) + e^{\mp   k \pi \bw / L}  \bar  T(\bw)\) \, ,
\ee
with  $ T(w)$ and $\bar T(\bar w)$ defined on the strip.  

Mapping the strip to the UHP via $z = e^{\frac{\pi}{L}w}$,  we can then write
\begin{align}\label{eq:DHamiltonian}
H_k = \frac{\pi}{L \cosh (2\theta) } \oint \left[ \cosh(2\theta) z - \frac{\sinh(2\theta)}{2} (z^{-k+1} +z^{k+1}) \right] T(z) - \frac{c\pi}{24L}\,,
\end{align}
where the contour integral is in the full complex plane  (see appendix \ref{App:MIM} for details). 

This makes it explicit that $H_k^{\pm}$ can be expressed in terms of Virasoro generators in the complex plane, giving 
\be
 H^{+}_k + H^{-}_k =  \frac{\pi}{L} \oint \(  z^{k+1} T(z)  + z^{-k+1} T(z) \)  = \frac{\pi}{L}  \(L_k+ L_{-k}\)\, .
\ee
Employing the map
\be
\tilde z^k =   -\frac{z^k \cosh  \theta  - \sinh  \theta }{z^k \sinh \theta    - \cosh \theta }  \,,
\ee
allows to bring the total Hamiltonian into the form
\be
\begin{aligned}\label{eq:deformedH}
H_k 
&= \frac{\pi}{L} \left[ \frac{1}{\cosh(2\theta)} \oint \tilde z \,   T(\tilde z)  - \frac{ c}{24}\left( 1- \frac{k^2-1}{\cosh(2\theta)}\right)\right]  \\
&=\frac{\pi}{L} \left[ \frac{1}{ \cosh(2\theta)} \tilde L_{0}  - \frac{c }{ 24}\left( 1- \frac{k^2-1}{\cosh(2\theta)}\right)\right] \,,
\end{aligned}
\ee
where $ \tilde L_{0}$ indicates the dilatation generator in the $\tilde z$-coordinate. Therefore, for  a primary $\O$
\be
e^{H_k \tE} \O(\tilde z, \bar {\tilde  z}) e^{- H_k \tE}  =   \lambda^{2 h_\O}  \O (\lambda \tilde z  , \lambda  \bar {\tilde  z})\, .
\ee
with  
\be
\lambda =\exp {\frac{\pi \tE}{L \cosh(2\theta)}}\, .
\ee
Mapping back to the UHP, the Euclidean evolution of a CFT primary operator $\O$ under the deformed Hamiltonian can be expressed using \eqref{eq:HeisO} with the coordinate transformation  
\be \label{eq:kfoldsol}
z_\tE^k = \frac{ \left[(1-\lambda^k ) \cosh (2 \theta )-(1+ \lambda^k )\right] z^k - (1-\lambda^k ) \sinh (2 \theta)}{ (1-\lambda^k )  \sinh (2 \theta )z^k  -(1-\lambda^k) \cosh (2 \theta )-(1+ \lambda^k)} \, .
\ee 
This can be seen as a $SL(2,\mathbb{R})$ transformation in the $k$-covering space of the original coordinate $z$ and generalizes the result for the Hamiltonians discussed in the previous section, which is readily recovered for $k=2$.

 \subsection{Entanglement entropy after the quench}

The time evolution of the entanglement entropy after a quench to the generally deformed Hamiltonian $H_k$ is  obtained with the analytic continuation to $n=1$ of the R\'enyi calculation \eqref{eq:Sn}
\be \label{eq:Snbis}
S^{(n)}_\sigma = \frac{1}{1-n}\log \Tr \rho_\sigma^n =  \frac{1}{1-n}\log    \bra{\psi_{AB}}  e^{H_{k} \tE} \Phi_n(w=i \sigma,\bar w=- i\sigma) e^{-H_{k} \tE}    \ket{\psi_{AB}}  \, .
\ee
Employing $z_\tE$ from \eqref{eq:kfoldsol} in the evolution equation \eqref{eq:HeisO}, one implements the time dependence as 
 \begin{align}  \label{eq:3ptgeneral} 
  \bra{\psi_{AB}}  e^{H_{k} t_E} &\Phi_n(w ,\bar w ) e^{-H_{k} t_E}    \ket{\psi_{AB}}  \no\\
  &= \( \frac{\del z_{t_E}}{\del w} \)^{h_n}   \( \frac{\del \bz_{t_E}}{\del \bw} \)^{h_n} \langle  \psi_{AB} (\infty)    \Phi_n (z_{t_E},\bar z_{t_E})   \psi_{BA}(0) \rangle_{\rm UHP}\,   .  
 \end{align}
 %
 
 \subsubsection{Evaluation of the time dependent correlator}

\begin{figure}[t!]
\centering
  \includegraphics[width=0.33 \textwidth]{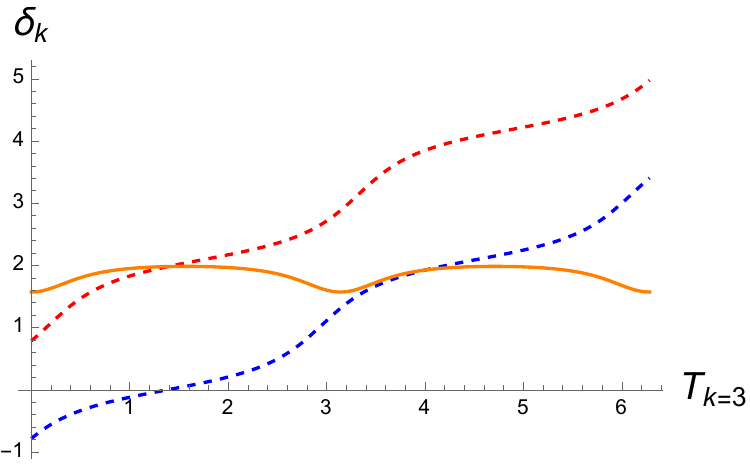} \hfill \includegraphics[width=0.33\textwidth]{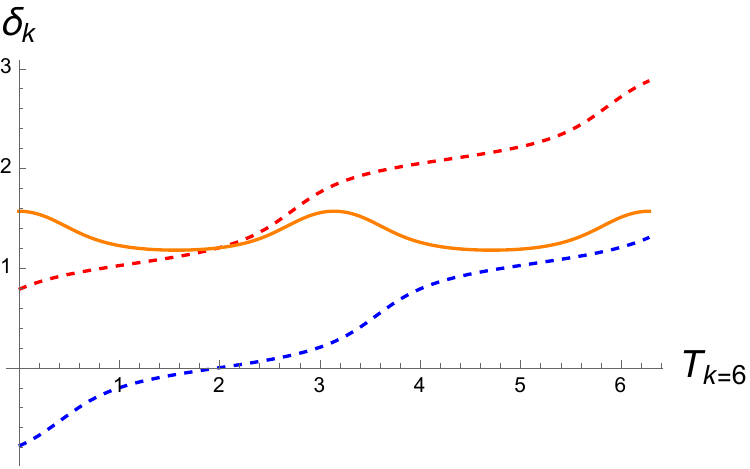}\hfill
  \includegraphics[width=0.33\textwidth]{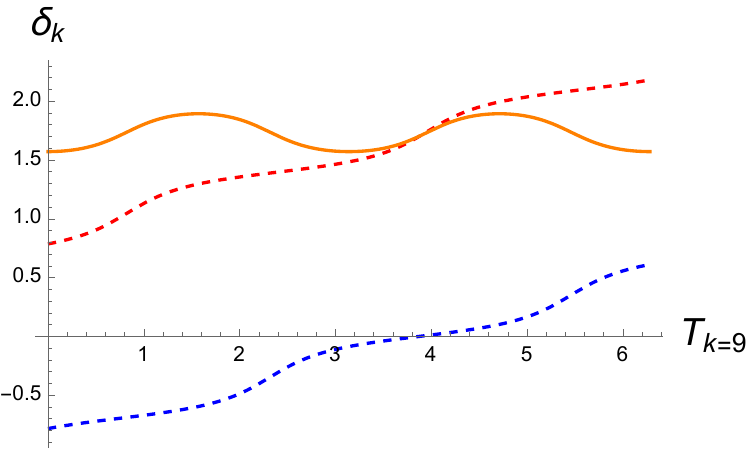}
\caption{Real time evolution of the phases of $z_t$ dashed red, $\bz_t$ dashed blue, and $\delta_k$. From left to right, $k=3,6,9$. In all plots $\theta=0.4$, $\sigma/L = \frac{1}{5}$.}
\label{fig:phasesGH}
\end{figure}

Continuing \eqref{eq:kfoldsol}  to real time $t_E \to it$ gives
\be \label{eq:ztimegen}
z^k_t= \frac{ \( \cos (T_{k}) \cos \left(\frac{k \pi \sigma }{2 L}\right)-e^{2 \theta } \sin (T_{k}) \sin \left(\frac{k \pi \sigma }{2 L}\right)+i \left(e^{-2 \theta } \cos \left(\frac{k \pi \sigma }{2 L}\right) \sin (T_{k})+\cos (T_{k}) \sin \left(\frac{k \pi \sigma  }{ 2L}\right)\right)\)^2}{{\cos ^2(T_{k})-\sin (2 T_{k}) \sin \left(\frac{ k \pi \sigma }{L}\right) \sinh (2 \theta )+\sin ^2(T_{k})   \left(\cosh (4 \theta )-\cos \left(\frac{k \pi  \sigma }{L}\right) \sinh (4 \theta )\right)}}
\ee
with  $\bz_t$ obtained via the replacement $\sigma \to - \sigma$. The rescaled time variable $T_k$ is defined as 
\be
T_{k} \equiv \frac{k\pi    t }{2L\cosh(2 \theta )} = \frac{k T_\theta}{2}\,.
\ee
The Lorentzian insertions $z_{t} $ and $\bar { z}_{t}$  start from their respective initial values, $ z_{t=0}  = e^{i \frac{\pi \sigma}{L}}$ and $\bar { z}_{t=0} = e^{-i\frac{\pi \sigma}{L}}$,  and move on the unit circle with phases increasing in time.  They go around the circle once over a rescaled  $T_{k} $  period   $k\pi$ (see fig.~\ref{fig:phasesGH}).

\begin{figure}[t!]
\centering
  \includegraphics[width=0.45 \textwidth]{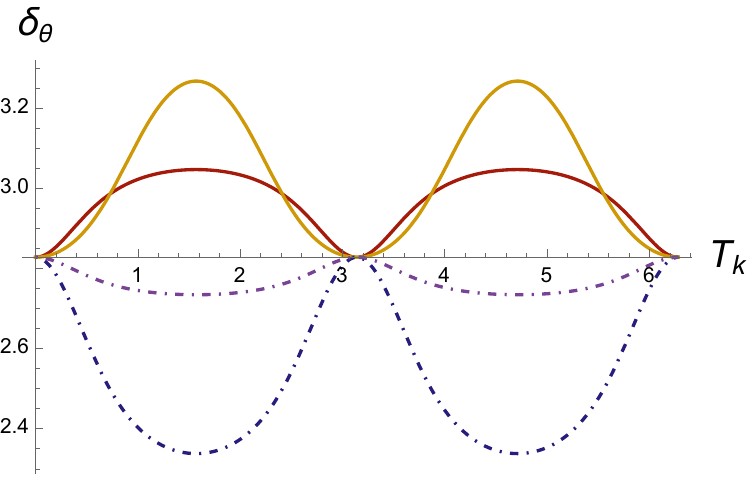}   \hfill
   \includegraphics[width=0.45 \textwidth]{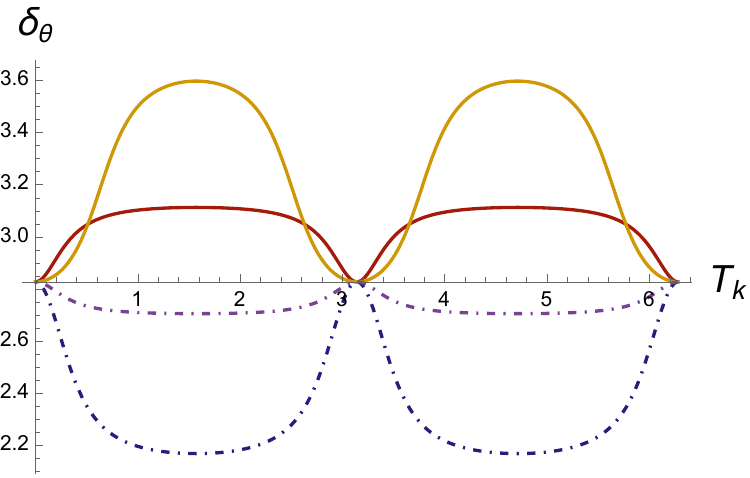} 
 \caption{Evolution of the phase $\delta_k$ with   $\sigma/L=0.45$. The solid lines are for $k=2,5$ going from the bottom-up at  $ \delta_{k,\text{ext}} $ which is a maximum, \ie, $ \frac{2m}{k} < \frac{\sigma}{L} < \frac{2m+1}{k}$.  The dashed-dotted lines are  for   $\delta_{k,\text{ext}}$ corresponding to a minimum, \ie, $ \frac{2m+1}{k} < \frac{\sigma}{L} < \frac{2(m+1)}{k}$, with $k=3,7$ from the bottom-up.   $\theta = 0.3$ on the left and $\theta = 0.6$ on the right, showing how the deformation affects the amplitude of the oscillations. }\label{fig:kdepGH}
\end{figure}

Using \eqref{eq:ztimegen} and evaluating the UHP correlator in \eqref{eq:3ptgeneral}  as the identity Virasoro block contribution in channel $A$ or $B$ gives    
\begin{align}
 \label{eq:AB3ptquenchgeneral}
  \bra{\psi_{AB}}  e^{H_{k} t_E} \Phi_n(w ,\bar w ) &e^{-H_{k} t_E}    \ket{\psi_{AB}}    =  \( \frac{\alpha \pi }{2 L}\)^{2h_n}   \(     f_k(T_{k} )^2 +   \sin ^2   \( \frac{k \pi  \sigma}{L}   \)   \)^{-h_n}   \\
&\times \max \left\{   A_{\Phi_n}       \left[     \sin^2 \( \frac{\alpha}{2} \delta_{k}(T_{k}) \)   \right]^{-h_n}  ; \, B_{\Phi_n}       \left[     \sin^2 \( \frac{\alpha}{2}( 2 \pi - \delta_{k}(T_{k})   ) \)   \right]^{-h_n}   \right\} \, . \nonumber 
\end{align}
 This is the analogue of  \eqref{eq:AB3ptquench} and  \eqref{eq:AB3ptquenchM} for the SSD and M\"obius cases.  
 In particular, $f_k$ is a generalization of the function $f_\theta$ defined in equation \eqref{eq:ftheta} for the M\"obius   Hamiltonian\footnote{Notice that $f_k$ also depends on the deformation  parameter $\theta$.}
\be \label{eq:fk}
f_{k}(T_k) \equiv  -  \sin ^2( T_{k}) \sinh(4 \theta ) +  \left(\cos ^2(T_k)+ \cosh (4 \theta ) \sin^2(T_k)\right)  \cos \( \frac{k\pi \sigma}{L} \) \,. 
\ee
 $ \delta_{k}$  is the phase difference between $z_t$ and $\bar z_t$, which can be expressed as
\be  \label{eq:phasek}
e^{i   \delta_{k}} \equiv \frac{z_t}{\bz_t}  = \( \frac{  \( f_k(T_{k} ) +  i  \sin    \( \frac{k \pi  \sigma}{L}   \)\)^2 }{  f_k(T_{k} )^2 +   \sin ^2   \( \frac{k \pi  \sigma}{L}   \)}
\)^{\frac{1}{k}} \, . 
\ee
As displayed in fig.~\ref{fig:phasesGH}, $\delta_k(T_k)$  oscillates with periodicity $\pi$ between the initial value $ \frac{2  \pi \sigma}{L}$  and the extremal value  
\be \label{eq:extremeK}
\delta_{k,\text{ext}}  = \frac{2 \pi \sigma}{L} +\frac{4}{k} \arctan\left( \frac{\sin \( \frac{k\pi \sigma}{L} \)}{\coth (2 \theta )-\cos \( \frac{k\pi \sigma}{L} \)  } \right)  \,.
\ee
$\delta_{k,\text{ext}}$ represents a maximum or  a minimum depending on the initial value, \ie, depending on the size  $\sigma$ of the entangling region. For   $\sin\( \frac{k\pi \sigma}{L} \)$ positive (negative), $\delta_{k,\text{ext}}$ is  the maximum (minimum) value reached at half periods of the effective time   $ T_{k}  = \pi/2,3\pi/2, \dots$\, . A few cases for different values of $k$   are illustrated in fig.~\ref{fig:kdepGH}.
In the M\"obius case $(k=2)$ the initial value was always the minimum (maximum) value for $\sigma$ smaller (larger) than half of the system (see fig.~\ref{fig:thetadepM}).  Here instead the extra parameter $k$ allows for both cases for any value of $\sigma$.  

We briefly comment on the equivalent of the SSD limit,  $\theta\to \infty$. Also here, in the limit, the returning phase which in the M\"obius case distinguished the  finite $\theta$  deformation from the SSD case disappears. 
 In particular the phase $\delta_k$  interpolates monotonically between the initial value and  $\delta_{k,\text{ext}}$ in  \eqref{eq:extremeK}  in the limit  $ \theta\to \infty$. This is attained asymptotically for $t \to \infty$ and reads  
\be
 \delta_{k,\text{ext}} =  2\pi \frac{(2m+1)}{k} ~~~ \text{where $m$ is such that}~~~ \frac{2m}{k}\le \frac{\sigma}{L} < \frac{2(m+1)}{k} \, . 
 \ee
The $k=2$  case corresponds to the M\"obius and SSD case. There $m=0$ for any $\sigma$, giving $\delta_{2,\text{ext}} = \pi$, consistently with the $t \to \infty$ limit of the phase in the SSD case discussed in section \ref{sec:Mobius}.

%
\subsubsection{Boundary-induced phase transition}\label{sec:GeneralH}
 
\begin{figure}[t!]
\centering
  \includegraphics[width=0.3 \textwidth]{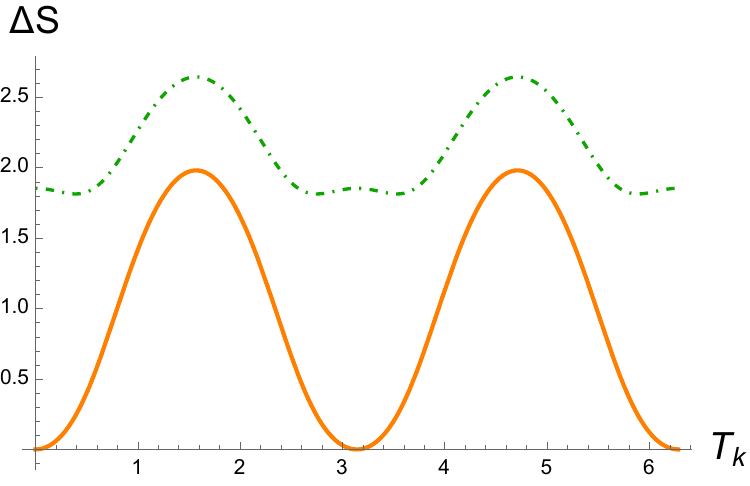} \hfill \includegraphics[width=0.3\textwidth]{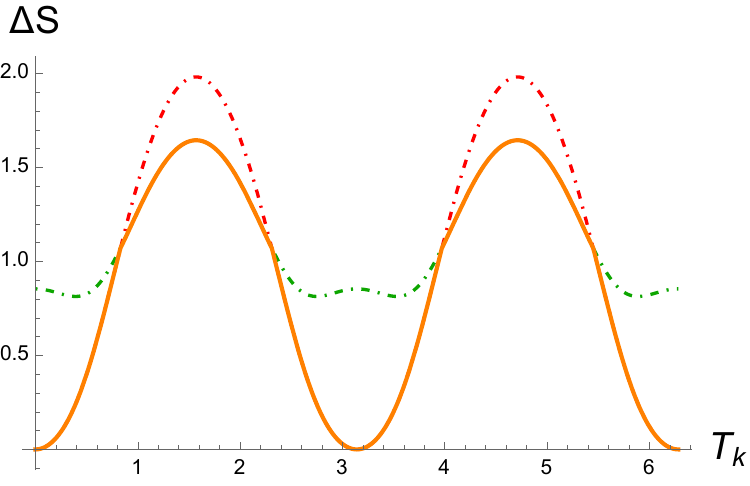} \hfill \includegraphics[width=0.3\textwidth]{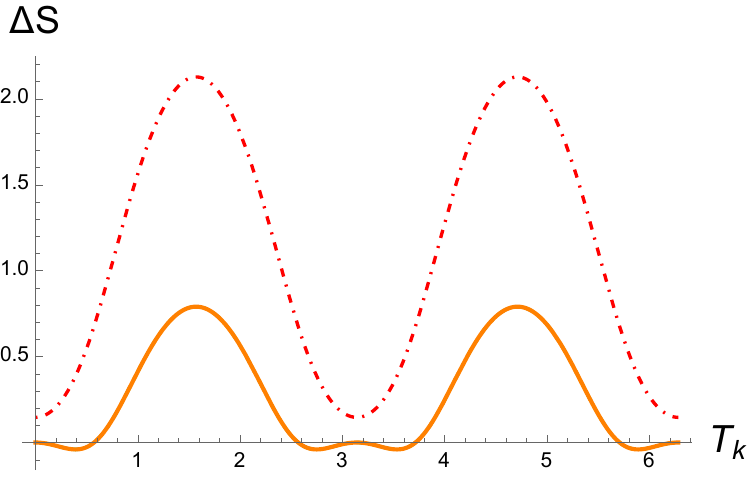}\\  
    \includegraphics[width=0.3 \textwidth]{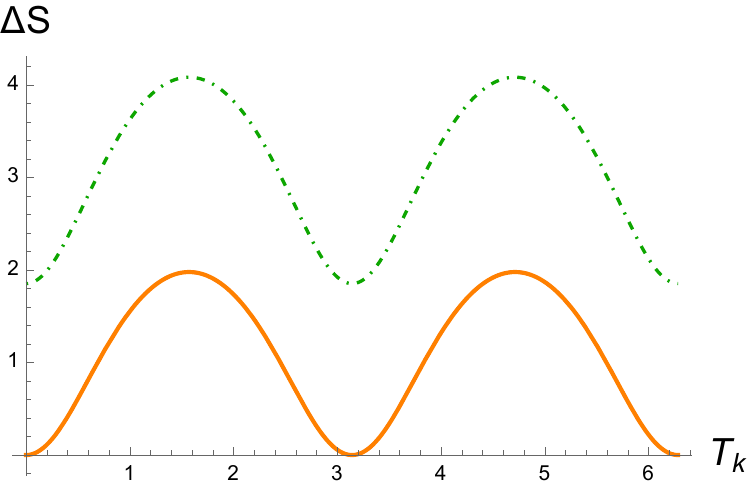} \hfill \includegraphics[width=0.3\textwidth]{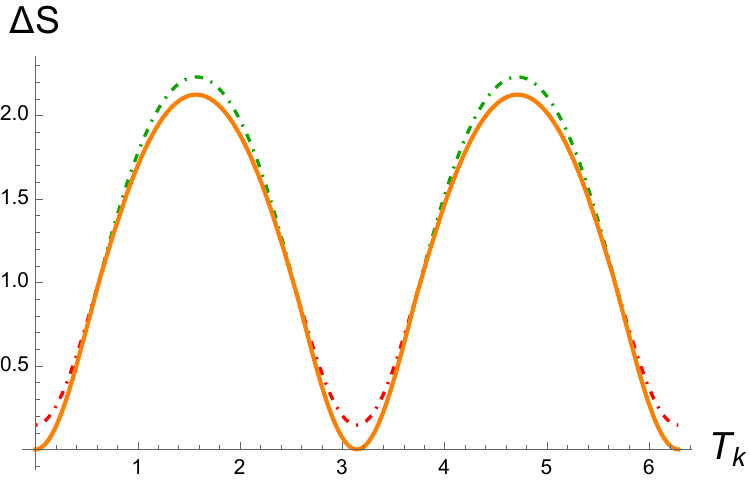} \hfill \includegraphics[width=0.3\textwidth]{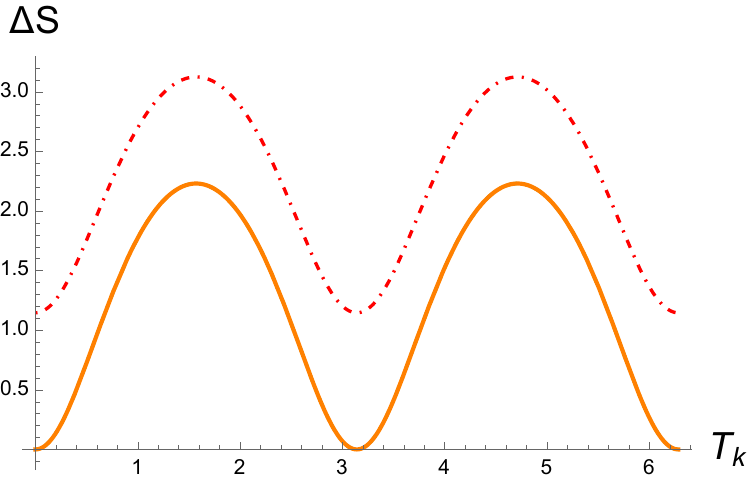}\\  
 \caption{Plot of the entanglement entropy evolution,  rescaled by $12/c$  with the initial value subtracted. 
The solid line is the entanglement entropy obtained from  \eqref{eq:CFTresultk}.  The dot-dashed lines correspond to channel $A$ (red) and $B$ (green) when in their non-dominant phase. In all plots $\sigma/L= 0.15,\alpha =0.5$ and $\theta =0.3$.  Top: $k=3$ and it corresponds to the case $\frac{2m}{k}<\frac{\sigma}{L}<\frac{2m+1}{k} $, with $m=0$. From left to right $\frac{12}{c}(s_B-s_A)=-1,-2,-3$. 
Bottom:  $k=8$ and this situation corresponds to the case $\frac{2m+1}{k}<\frac{\sigma}{L}<\frac{2(m+1)}{k} $, with $m=0$. From  left to right $\frac{12}{c}(s_B-s_A)=-2,-3,-4$. }\label{fig:EEk}
\end{figure}

The expression for the entanglement entropy follows in a straightforward manner using the result \eqref{eq:AB3ptquenchgeneral} in \eqref{eq:Snbis} and continuing to $n=1$
\begin{align}
 \label{eq:CFTresultk}
S(T)= \frac{c}{12}  \log &  \[ \(  \frac{2 L}{ \pi \alpha \epsilon } \)^2 \(     f_k(T_k)^2+   \sin ^2   \( \frac{k \pi  \sigma}{L}  \)  \)  \]\\
&+\min \left\{
 \frac{c}{12}  \log  \sin^2\(\frac{\alpha}{2}    \delta_k (T_k) \)  + s_A;  \frac{c}{12}  \log   \sin^2\(\frac{\alpha}{2} \(2 \pi   -    \delta_k (T_k)  \)  \) + s_B\right\} \,  .\nonumber
\end{align}
$f_k(T_k)$ is defined in equation \eqref{eq:fk} and $\delta_k$ oscillates between the initial value $ \delta_k(T_k=0) = 2 \pi \sigma /L$ and the extremal value $\delta_{k,\text{ext}} $ in \eqref{eq:extremeK}. Minimizing between the two channels at each instant of time gives the entanglement entropy evolution. Figure~\ref{fig:EEk} illustrate some examples for different choices of the parameters. 

Similarly to the M\"obius quench, a dynamical transition between the two channels can happen only for specific ranges of the boundary entropies. This condition can be summarized as
\be \label{eq:transcondk}
\begin{aligned}
&\frac{c}{6}\log \frac{\sin  \( \frac{\alpha\pi \sigma}{L} \)}{ \sin \( \frac{\alpha\pi (L-\sigma)}{L} \)} < s_B-s_A <  \frac{c}{6} \log \frac{\sin \( \frac{\alpha \delta_{k,\text{ext}}}{2}\)}{ \sin \(\frac{\alpha (2\pi- \delta_{k,\text{ext}} )}{2} \)} ~ & \text{for}\quad &  \frac{2m}{k} < \frac{\sigma}{L} < \frac{2m+1}{k} \, , \\  
&\frac{c}{6}\log \frac{\sin \( \frac{\alpha \delta_{k,\text{ext}}}{2}\)}{ \sin \(\frac{\alpha (2\pi- \delta_{k,\text{ext}} )}{2} \)} <   s_B-s_A <  \frac{c}{6} \log \frac{\sin  \( \frac{\alpha\pi \sigma}{L} \)}{ \sin \( \frac{\alpha\pi (L-\sigma)}{L} \)} ~ & \text{for}\quad &     \frac{2m+1}{k} < \frac{\sigma}{L} < \frac{2(m+1)}{k}
\end{aligned}
\ee
with $m = 0,1,2,\dots,[\frac{k}{2}-1]$,  with $[\cdot]$ indicating the integer part. 

For a given set of parameters,  when $s_B-s_A$ falls in the above range one observes a phase transition between the entanglement evaluated in the two channels at each time $T_{k}^*$ such that  
 \be   \label{eq:PhTr}
\frac{c}{6} \log  \frac{ \sin\(\frac{\alpha}{2} \delta_k(T_{k}^*)   \)}{ \sin\(\frac{\alpha}{2} \( 2\pi - \delta_k(T_{k}^*)\)    \)}  = s_B-s_A \, .
\ee
The periodic behavior induced by the M\"obius type deformation at finite $\theta$  implies that the  transition happens twice during a period of $T_k$. One goes from the initial channel to the other at some time $T_{k}^*$ and then back to the initial channel at $\pi- T_{k}^*$. 

When the difference of  boundary entropies is outside the range \eqref{eq:transcondk}, one channel always dominates over the other. More precisely, when $\frac{2m}{k}<\frac{\sigma}{L}<\frac{2m+1}{k} $, the dominant channels are
\be
\begin{aligned}
A\text{-channel}&~~\text{for} ~~ s_B - s_A \ge  \frac{c}{6}  \log \frac{\sin \( \frac{\alpha \delta_{k,\text{ext}}}{2}\)}{ \sin \(\frac{\alpha (2\pi- \delta_{k,\text{ext}} )}{2} \)}  \,,\\
B\text{-channel}&~~\text{for} ~~ s_B - s_A \le \frac{c}{6} \log \frac{\sin  \( \frac{\alpha\pi \sigma}{L} \)}{ \sin \( \frac{\alpha\pi (L-\sigma)}{L} \)} \, .
\end{aligned}
\ee
In the complementary range $ \frac{2m+1}{k} < \frac{\sigma}{L} < \frac{2(m+1)}{k}$, $A$ and $B$ simply get  exchanged
\be
\begin{aligned}
B\text{-channel}&~~\text{for} ~~ s_B - s_A \le \frac{c}{6}   \log \frac{\sin \( \frac{\alpha \delta_{k,\text{ext}}}{2}\)}{ \sin \(\frac{\alpha (2\pi- \delta_{k,\text{ext}} )}{2} \)}  \,,\\
A\text{-channel}&~~\text{for} ~~ s_B - s_A \ge \frac{c}{6}  \log \frac{\sin  \( \frac{\alpha\pi \sigma}{L} \)}{ \sin \( \frac{\alpha\pi (L-\sigma)}{L} \)} \, .
\end{aligned}
\ee
%
 
\section{Holographic dual}\label{sec:holography}

In this section we present the gravity dual to the BCFT setup with distinct boundary conditions of section~\ref{sec:BCCop} and evaluate the holographic entanglement entropy. We then show how to reproduce holographically the result for the time dependent entanglement entropy in the post quench state. 

Discussions of aspects of the holographic dual of the BCFT setup and related AdS/BCFT constructions appeared in \cite{Geng:2021iyq,Miyaji:2022dna,Biswas:2022xfw}.  
Here we will closely follow the approach of \cite{Miyaji:2022dna}, while modifying their construction as to adapt it to a BCFT on an interval at zero temperature.

\subsection{Gravitational action and equations of motion}

\begin{figure}[t!]
\centering
         \includegraphics[width=0.5 \textwidth]{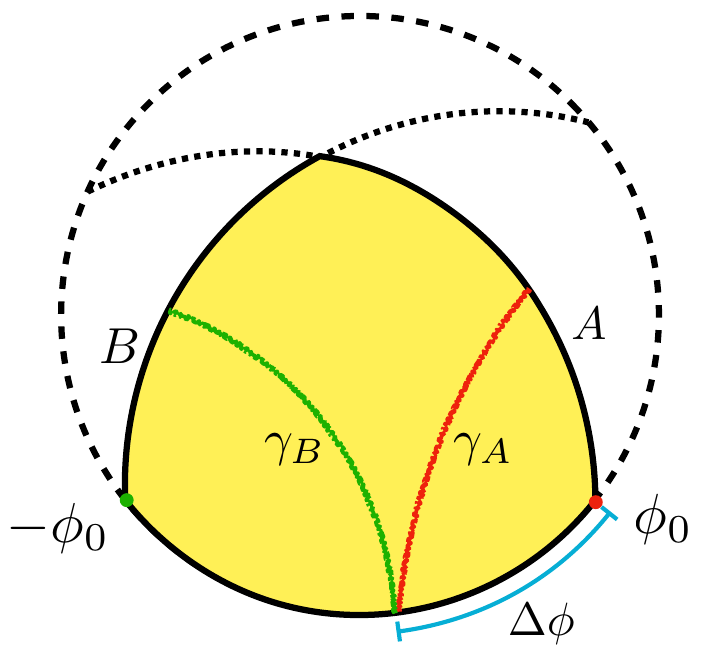}
\caption{On a given Cauchy slice, the two EOW branes corresponding to the two conformal boundary conditions imposed in the CFT anchor at $\phi_0$ and $-\phi_0$  on the asymptotic boundary and intersect in the bulk.  The shaded  region is the gravitational dual to the CFT  we are considering. A defect in the bulk supports the intersection of the two branes. For a given boundary interval  $\Delta \phi$ there will be in general two geodesic configurations $\gamma_{A}$ and  $\gamma_{B}$ emanating from the asymptotic boundary and anchoring on the  corresponding EOW brane. }
\label{fig:interbranes}
\end{figure}

The holographic description of the BCFT setup of section~\ref{sec:BCCop}  involves two end of the world (EOW) branes with different tensions, intersecting at a defect in the bulk of global AdS$_3$. The branes single out a region of  the AdS$_3$ spacetime which provides the dual geometry for  the BCFT on an interval with mixed boundary  conditions  (see fig.~\ref{fig:interbranes}).

The total gravitational action of the holographic dual is given by \cite{Miyaji:2022dna}
\be
I = I_{\rm EH} + \sum_{i =A,B} I_{\rm EOW, i} + I_{\rm DEFECT} + I_{\rm CT}\,.
\ee
The first contribution is the Einstein-Hilbert action with negative cosmological constant $\Lambda$ plus a Gibbons-Hawking-York term at the AdS boundary $\del$AdS 
\be
I_{\rm EH} = - \frac{1}{16 \pi G_N}\int_{\rm AdS} \sqrt{- g} \(R - 2 \Lambda \) - \frac{1}{8 \pi G_N} \int_{\del {\rm AdS}} \sqrt{- \gamma} K\,.
\ee

 To implement holographically the two distinct boundary conditions of the BCFT, we add in the AdS$_3$ bulk two EOW branes $\Sigma_i$.\footnote{While strictly speaking these objects are strings in AdS$_3$, we will use the nomenclature that is most common in this literature and refer to them as branes, as in the higher dimensional settings.} The anchoring points on the boundary $\del$AdS corresponds to the BCFT interval endpoints.
 For purely tensional branes, the action is
\be \label{eq:IEOW}
 I_{\rm EOW, i} = \frac{1}{8 \pi G_N} \int_{\Sigma_i} \sqrt{- h} T_i - \frac{1}{8 \pi G_N} \int_{\Sigma_i} \sqrt{- h} K  - \frac{1}{8 \pi G_N} \int_{\del {\rm AdS} \cap \Sigma_i} \sqrt{- \delta} \( \pi - \Theta_i \)\, .
\ee
Here $h_{\a\b}$ is the induced metric on the brane and $K_{\a\b}$ its extrinsic curvature. For each brane  the equations of motion read
\be \label{eq:israel}
K_{\a\b} = (K - T_i) h_{\a\b} \, .
\ee
This sets the shape of the EOW brane $\Sigma_i$ in terms of its tension $T_i$ and boundary conditions. 

The last term in  \eqref{eq:IEOW} is a Hayward term at the corner $\del {\rm AdS} \cap \Sigma_i$ between the AdS boundary and each brane $\Sigma_i$. $\delta$ is the induced metric and  the internal angle $\Theta_i=\pi$ on shell. 

The two branes have in general different tensions $T_i$ and will intersect in the bulk on a worldline $\Gamma$. There one needs to include an interaction term in the form of a defect to hold together two distinct branes. The simplest such action is a slight modification of a Hayward term given by
\be
I_{\rm DEFECT} =  - \frac{1}{16 \pi G_N}\int_\Gamma \sqrt{- h_\Gamma} \(\theta_0 - \hat \theta \)\, .
\ee
 $\theta_0 - \pi$ can be regarded as the effective tension of the defect and $\hat\theta$ is the internal angle at the branes intersection. On shell this is related to the defect tension via the condition
\be \label{eq:intersangle}
\hat \theta = \theta_0 \,  .
\ee

Finally, the action includes appropriate counterterms $ I_{\rm CT}$ to renormalize the divergences coming from near the AdS boundary \cite{Miyaji:2022dna}.

\subsection{Intersecting branes in AdS$_3$}

We first briefly review the AdS dual to a BCFT on an interval with equal boundary conditions at the endpoints.

Consider empty AdS$_3$ with  metric
\be \label{eq:AdS3metric}
ds^2 =- \frac{r^2 + 1}{R^2}dt^2 + \frac{dr^2}{ r^2 + 1} + r^2 d\phi^2 \,,
\ee
with $r \in [0, \infty)$, $t \in (- \infty, \infty)$ and $\phi \in [-\pi, \pi ]$. The AdS radius is set to one and $R$ denotes the radius of the boundary circle. 

Take a purely tensional brane of tension $T$ emanating from a point $0 \le \phi_0 < \pi$  on  $\del$AdS$_3$.  On shell,  the brane has the  following static profile 
\be \label{eq:bprof}
r = \frac{k}{\sin \(\phi - \phi_0\) }  \, , 
\ee
where we have defined the combination
\be
  -\infty <k \equiv \frac{ T}{\sqrt{1 - T^2}}< \infty \, . 
 \ee
The brane connects antipodal points on  $\del$AdS$_3$.  If  $k < 0 $   the brane spans in the bulk the angular interval  $-\pi + \phi_0 \le \phi \le \phi_0$ while for $k >0$ it spans the interval  $\phi_0 \le \phi \le \pi \cup - \pi  \le \phi \le -\pi + \phi_0$. 
Retaining the AdS bulk geometry comprised between the brane and the  portion of boundary with angular range $-\pi + \phi_0 \leq \phi \leq \phi_0$ gives the dual for the BCFT on the strip with equal boundary conditions (see for example \cite{Takayanagi:2011zk,Fujita:2011fp}).
 
To implement two distinct boundary conditions as in section~\ref{sec:BCCop}, we consider two static branes anchored at symmetric points about $\phi = 0$, \ie, at $\pm \phi_0$ with $0 < \phi_0< \frac{\pi}{2}$. In the intersecting configuration, we retain the AdS bulk portion that has angular boundary range $-\phi_0 \leq \phi \leq \phi_0$,
as illustrated in fig.~\ref{fig:interbranes}.\footnote{One could in principle also consider the complementary range $-\pi \le \phi \le -\phi_0 \cup \phi_0 \leq \phi \leq \pi$  or, equivalently,  $\frac{\pi}{2} < \phi_0< \pi$  \cite{Miyaji:2022dna}. }

We   parametrize the two branes profiles as
\be \label{eq:branesprofile}
r = \frac{k_A}{\sin \(\phi - \phi_0\) } \qquad \quad \text{and}\quad \qquad  r = - \frac{k_B}{\sin \(\phi+ \phi_0\) } \, ,
\ee
without making any assumption about the signs and values of the tension parameters. With this choice,  for equal tensions $k_A = k_B$ and $\phi_0 =  \pi/2$   the two branes coincide. 
Notice that for $k_A >0$ ($k_A < 0$) the first brane spans the angular coordinate range $\phi_0 \le \phi \le \pi \cup -\pi \le \phi \le -\pi + \phi_0$ ( $-\pi + \phi_0 \le \phi \le \phi_0$). If $k_B >0$ ($k_B < 0$) the second brane spans the angular coordinate range  $ -\pi \le \phi \le - \phi_0 \cup  - \phi_0 + \pi \le \phi \le  \pi $ ($- \phi_0 \le \phi \le - \phi_0 + \pi $) (see fig.~\ref{fig:interbranes}).  

On any fixed time slice, the branes intersect at a point  $(r_*, \phi_* )$ such that  
\bea \label{eq:intersection}
r_*^2 &=& \frac{k_A^2 + k_B^2 + 2 k_A k_B \cos 2 \phi_0}{\sin^2 2 \phi_0} \label{eq:rintersection} \\
\tan \phi_*  &=& \frac{k_B - k_A}{k_A+ k_B} \tan \phi_0 \, . \label{eq:phiintersection}
\eea
In particular we have  
\bea
\phi_* \in 
\left\{\begin{array}{cl}
\left[\phi_0 , \pi - \phi_0\right] &\mbox{if}~~k_A>0, k_B < 0 \\
\left[\pi -\phi_0 , \pi\right] \cup \left[ -\pi , -\pi + \phi_0 \right] &\mbox{if}~~k_A > 0, k_B > 0 \\
 \left[-\phi_0 , \phi_0\right] &\mbox{if}~~k_A < 0, k_B < 0 \\
\left[-\pi + \phi_0 , -\phi_0 \right] &\mbox{if}~~k_A <0, k_B > 0 
\end{array}\right. .
\eea
Also, notice that in the limit $\phi_0 \to \pi/2$, the radial coordinate $r_*$ of the branes intersection goes to infinity, but this limiting value can be approached for any choice of  $T_A$ and  $T_B$.

The angle $\hat\theta$ between the branes at the intersection point is obtained in terms of the unit normals $n_A, n_B$. 
Defining them to be outward-pointing with respect to the bulk region between the branes gives
\be
\begin{aligned}
 n_{A \mu} &=\pm \sqrt{\frac{\sin^2 (\phi - \phi_0)}{1+ k_A^2}} \(0,1, \frac{k_A}{\sin (\phi - \phi_0) \tan(\phi -\phi_0)} \)\,,\\
 n_{B \mu} &= \pm \sqrt{\frac{\sin^2 (\phi + \phi_0)}{1+ k_B^2}} \(0,1, - \frac{k_B}{\sin (\phi + \phi_0) \tan(\phi + \phi_0)} \) \,,  
\end{aligned}
\ee
with the $\pm$ sign respectively for positive and negative values of the tension parameters. 
Keeping track of the sign of the tensions and using  equation~\eqref{eq:branesprofile}, \eqref{eq:rintersection} and \eqref{eq:phiintersection}, one obtains the relation 
\be
\cos (\pi -\hat\theta)  = n_{A \mu} n_{B \nu} g^{\mu \nu} =  \frac{k_A k_B- \cos 2 \phi_0}{\sqrt{(1+k_A^2)(1+ k_B^2)}}\, 
\ee
for any sign of $k_A$ and $k_B$. 

On shell, the brane intersection angle coincides with the tension parameter of the defect sustaining the intersection,  $\theta_0 =\hat \theta$,  which is then related to the  opening angle $\phi_0$ as
\be \label{eq:braneangle}
\cos 2\phi_0 = k_A k_B + \sqrt{(1+k_A^2)(1+ k_B^2)} \cos \theta_0\, .
\ee
The case in which the branes have the same tension and there is no intersection nor defect, $ \theta_0 = \pi$, yields $\phi_0 = \pi/2$ consistently with the parametrization \eqref{eq:branesprofile}.

\subsubsection{Matching the BCFT setup}

We now match the BCFT parameters of the setup in section~\ref{sec:BCCop}  to the holographic ones. 
The length $L$ of the CFT interval is given  by the opening between the two branes at the AdS boundary
\be \label{eq:sizeL}
L = 2 \phi_0 R \,.
\ee

The conformal dimension $h_\psi$ of the BCC operator is directly related to the opening angle $\phi_0$. The relation is obtained remembering that $h_\psi$ is determined by the gap in energy between the state with and without BCC  operator, \ie,  between the state   with mixed boundary conditions and the ground state with homogeneous boundary conditions (see  equation  \eqref{hBCC}). Holographically this is the energy difference between the configuration with two intersecting branes and  the one with a single brane.\footnote{This is actually independent from the tension of the brane, \eg, \cite{Grimaldi:2022suv}.} Evaluating the corresponding bulk actions, the shift in energy is found to be \cite{Miyaji:2022dna} 
\be
h_\psi  =\frac{L\,  \Delta E}{\pi}  =  \frac{c}{ 24  } \(1 -  \frac{  4 \phi_0^2 \, }{ \pi^2  } \)
\ee 
giving  
\be\label{eq:BCCweight}
\frac{2 \phi_0}{\pi} =\sqrt{ 1 - \frac{24 h_\psi}{ c}} 
\,. 
\ee

Notice that $h_\psi\propto c$ and the spectrum goes from arbitrarily close to $0$ to $c/24$ \cite{Miyaji:2022dna,Biswas:2022xfw}. Therefore, in  the holographic realization there exists a BCC operator  implementing the change in boundary conditions with an arbitrary small change in energy. 
That is, for any set of mixed  boundary conditions holographically specified by $T_A$ and $T_B$,  the lowest energy state has at leading order in $c$ the same energy of a the ground state with homogeneous  boundary conditions.

\subsection{Holographic entanglement entropy}\label{sec:HEEdyn}

To reproduce holographically the entanglement entropy result for the strip with different  boundary conditions, equation~\eqref{eq:EEstatic}, we shall evaluate the Ryu-Takayanagi \cite{Ryu:2006bv} formula  
\be \label{eq:RT}
S = \frac{ \text{Area}\, \gamma}{4G_N}
\ee
in the static geometry with  EOW branes \cite{Takayanagi:2011zk,Fujita:2011fp}. In AdS$_3$, the surfaces $\gamma$  are regulated geodesics  and  the area term is evaluated as the geodesic proper length, $\text{Area} \, \gamma = d_\gamma$. 

In our setup we need to consider  the minimal length  geodesic extending from a  point on the boundary to the EOW branes. There are in general  two competing, locally extremal, curves one should take into account. The minimal length geodesic extending into the bulk and ending on the brane with tension parameter $k_A$ and the one ending on the brane with tension  $k_B$, see fig.~\ref{fig:interbranes}.

\subsubsection{Geodesic computation}
Let us start reviewing the case of a geodesic stretching between a boundary point and a single brane in AdS$_3$. 
The proper length $d_\gamma$ of a constant time geodesic $\gamma$ in AdS$_3$ between two generic points is given by
\be \label{length}
\cosh  d_\gamma  =\sqrt{ (r_1^2 + 1)(r_2^2 + 1) }-  r_1 r_2 \cos  \( \phi_1-\phi_2\)  \, .
\ee
 We take one extrema to be at a generic point $(r_b,\phi_b)$ on the EOW brane with profile
\be \label{eq:branesingle}
 r_b    = \frac{k}{\sin  \phi_b } \, .
 \ee
The other is taken on the regulated AdS boundary  at $r_{\infty}$ displaced by $\tilde\phi$ from the brane anchoring point at $\phi =0$, that is at $(r_\infty, \tilde \phi)$.
For the length   $d_\gamma$ this gives 
\be  \label{eq:distanceb}
\cosh  d_\gamma  \approx   r_{\infty}  \( \sqrt{     \frac{k^2}{\sin^2 \phi_b} + 1  }- \frac{k}{\sin \phi_b } \cos ( \tilde \phi - \phi_b  ) \) \, ,
\ee
where we made a large $r_{\infty}$ expansion and used  \eqref{eq:branesingle}  to express $r_b$ in  terms of $\phi_b$.

Extremizing the length over the possible ending points on the brane sets\footnote{Notice that for $\pi/2 <|  \tilde \phi|<\pi$ this means that $\pi/2< |\phi_b | < \pi$.}
\be \label{eq:prof0}
\sin \phi_{b,\rm ext}= \frac{k \sin |  \tilde \phi|}{\sqrt{k^2 + \cos^2   \tilde \phi}}\,  
\ee
and selects the geodesic with minimal length amongst all geodesics ending on the brane. This has length 
\be 
 d_\gamma  =  \log \[ 2 r_{\infty}  \( - k   \sin  \tilde  \phi  + \sqrt{ 1 + k^2 } \, \sin  |  \tilde \phi| \)\]  \, . \label{eq:dgeod}
\ee

The single brane analysis is immediately extended to the situation with two intersecting EOW branes as in equation \eqref{eq:branesprofile}. We parametrize  the boundary point from  where the geodesic emanates in terms of the distance  $0<\Delta \phi<2\phi_0$ from brane $A$. The point is then at the angular location $\phi_0 - \Delta \phi$.

There are now two locally minimal geodesics  to consider. 
One is the  geodesic $\gamma_A$ ending on the EOW brane $A$, which has length 
\be 
 d_{\gamma_A} =  \log \[ 2 r_{\infty}  \sin \Delta \phi  \]+ \arcsinh k_A   \, . \label{eq:lgA}
\ee
The other geodesic $\gamma_B$ is the minimal length one between those ending on the brane $B$. This  has length
\be 
 d_{\gamma_B} =  \log \[ 2 r_{\infty}  \sin(2\phi_0 -\Delta \phi)\] +\arcsinh k_B   \,  .\label{eq:lgB}
\ee
Notice  that this is obtained from \eqref{eq:dgeod} taking into account that the equation giving the brane profile $B$ in  \eqref{eq:branesprofile} has an overall extra minus sign as compared to \eqref{eq:branesingle}.  
 
For any given boundary interval $\Delta\phi$, the geodesic which has minimal length between \eqref{eq:lgA} and \eqref{eq:lgB} is the one selected by the Ryu-Takayanagi formula \eqref{eq:RT}
\be \label{eq:SRT}
S = \frac{\min\{ d_{\gamma_A} , d_{\gamma_B}\}}{4 G_N}\, .
\ee

As a function of $\Delta\phi$, a  dominance  transition from  one geodesic to the other happens for a value  $\Delta\phi_t$ corresponding to $ d_{\gamma_A}  =  d_{\gamma_B}$. Namely
\be
\cot \Delta \phi_t =\csc 2\phi_0 \,  \frac{k_A + \sqrt{1+ k^2_A} }{k_B + \sqrt{1+ k^2_B} }  +\cot  2 \phi_0  \,,
\ee
which is consistently within the range $0<\Delta\phi_t<2\phi_0$.

Notice that the two geodesics may not always exist at the same time in our setup.  For instance, $\gamma_A$ may cease to exist for a large enough  $\Delta \phi> \Delta \phi_A$. The limiting value corresponds to the situation where the endpoint of  $\gamma_A$ on the brane $\phi_{b,\rm ext}$  coincides with the branes intersection point $\phi_*$. That is, when
\be 
\sin (\phi_* - \phi_0)= \frac{k_A \sin \Delta \phi_A}{\sqrt{k_A^2 + \cos^2   \Delta \phi_A}}\,.
\ee
Substituting the explicit expression \eqref{eq:phiintersection}  for $\phi_*$   gives\footnote{This can formally give a value of $\Delta\phi_A$  outside the physical range $0< \Delta \phi < 2\phi_0$. In such a case $\gamma_A$ always exists in the setup we are looking at. A similar consideration applies to $\Delta\phi_B$.  }
\be
\cot \Delta\phi_A  = -\frac{\csc 2 \phi_0 \(k_A \cos 2   \phi_0 +k_B\)}{\sqrt{k_A^2+1}}  \, . 
\ee
Similarly  $\gamma_B$ may not exist for a small enough value of $\Delta \phi < \Delta \phi_B$. Here $\Delta \phi_B$ is given by
\be
\cot (2\phi_0-\Delta\phi_B)  = -\frac{\csc  2 \phi_0 \(k_B \cos (2\phi_0)+k_A\)}{\sqrt{k_B^2+1}}\,. 
\ee 
However, one can check explicitly that 
\be
\Delta\phi_B <  \Delta \phi_t < \Delta\phi_A \, , 
\ee
so that at the transition point both geodesic configurations do actually exist in our setup. 
 
\subsubsection{Matching the BCFT computation}

To show that the holographic entanglement entropy in \eqref{eq:SRT} matches the static result in section~\ref{sec:univSEE}, we need to express it in terms of the dual CFT parameters.  For the BCFT cutoff $\epsilon$ and  size of the entangling interval $\sigma$, we have the relations
\be
\frac{ r_\infty}{R} = \frac{1}{\epsilon}  
 \, \qquad \qquad R \Delta \phi = \sigma  \, . 
\ee 
These directly follow from matching the conformal boundary of AdS metric \eqref{eq:AdS3metric} with the  BCFT geometry.  Expressing everything in terms of the physical size $L$ through \eqref{eq:sizeL} and using  the relation \eqref{eq:BCCweight} yields
\be
r_\infty =  \frac{L }{\alpha \pi \epsilon}
 \, \qquad \qquad \Delta \phi = \frac{\sigma L}{\alpha \pi } \, ,
\ee 
where we have used $\alpha = \sqrt{ 1 - 24 h_\psi/ c}$ (see equation \eqref{eq:alphadef}).

Substituting these  into the expressions for the geodesics length  \eqref{eq:lgA} and \eqref{eq:lgB} and using the relation $c = {3 /2 G_N}$,   we get from  \eqref{eq:SRT}
\be \label{eq:HEEstatic}
S = \frac{c}{6}\log  \frac{2 L }{ \pi \alpha \epsilon} +  \text{min}  \left\{ \frac{c}{6} \log  \sin\frac{   \alpha \pi  \sigma}{L} +   \frac{c}{6} \, {\rm arcsinh} \, k_{A}   ; ~ \frac{c}{6} \log  \sin   \frac{ \alpha \pi (L - \sigma)}{L} +  \frac{c}{6} \, {\rm arcsinh} \, k_{B}   \right\} \, ,
\ee

By matching the brane ${A,B}$ with the conformal boundary condition $A,B$ in the CFT via the identification  \cite{Takayanagi:2011zk,Fujita:2011fp}
\be
s_{A,B} = \frac{c}{6} \, {\rm arcsinh} \, k_{A,B}  \, , \label{eq:sTrel}
\ee
 one exactly reproduces the result  \eqref{eq:EEstatic} for the static case  derived in section~\ref{sec:univSEE}.\\
  
The holographic computation of the entanglement entropy in the post quench state involves considering geodesics in a AdS$_3$ time dependent geometry  \cite{Ryu:2006bv,Hubeny:2007xt,Takayanagi:2011zk,Fujita:2011fp}. In principle it should be possible to work explicitly the form of this dual geometry, which one expects to be constructed from two EOW branes with a non-trivial time dependent profile.\footnote{See \cite{Goto:2021sqx,Goto:2023wai,Jiang:2024hgt} for this type of analysis performed in the case of a CFT on a circle.}  Here we will simply show how to reproduce the time dependent entanglement entropy.

The static state yielding the above result and the time dependent post quench state are related on the CFT side by a transformation of the conformal coordinates. 
This map can be extended to a diffeomorphism in AdS$_3$ that relates the geometries dual to the CFT states described in the two sets of coordinates  \cite{Banados:1998gg,Roberts:2012aq}.\footnote{A subtlety is that the time governing the evolution $\tE$ appears as a parameter in the map and not as a component of the complex CFT coordinates. The map relates for each  $\tE$ the initial state static geometry to an auxiliary geometry with its own time slicing where the geodesic length giving the time dependent result can be computed.  
See, \eg, \cite{Caputa:2018kdj,Erdmenger:2021wzc} for related discussions.}
The length of geodesics relevant for the post quench state can then be obtained using this local equivalence of  AdS$_3$ geometries (see, \eg, \cite{Hartman:2013qma,Asplund:2014coa,Asplund:2013zba}). 

Let us start expressing the static result \eqref{eq:HEEstatic} using coordinates that match those defined on the strip in section \ref{sec:strip}. For this we can simply focus on the asymptotic form of the AdS$_3$ metric \eqref{eq:AdS3metric}, which we can write in Euclidean as
\be
ds^2 \approx \frac{dr^2}{r^2} + \frac{r^2}{R^2} ( d \tau^2 + R^2 d\phi^2) = \frac{dw d\bw + du^2}{u^2} \ . 
\ee
We made the identifications
\be
u  =\frac{R}{r}  \qquad   w =\tau + i R \phi \qquad  \bw = \tau - i R \phi\, .
\ee
and $(w,\bw)$ match the Euclidean coordinates on the strip.

The entanglement entropy result obtained from the static geometry analysis is then rewritten as
\be \label{eq:Sholostatic}
S =   \text{min} \left \{ \frac{c}{6} \log  \frac{ 2L }{ \pi \alpha u_\infty} \sin\frac{   \alpha \pi  |w_\infty-\bw_\infty|}{2 L} +  s_A ; ~ \frac{c}{6} \log  \frac{ 2L }{ \pi \alpha u_\infty} \sin   \frac{ \alpha \pi (2L -  |w_\infty-\bw_\infty|)}{2L} + s_B \right\} \, 
\ee
and using the identifications  
\be
\begin{aligned} \label{eq:ident}
w_\infty &=  i \sigma  \\ 
\bar w_\infty & =  - i \sigma   \\ 
u_\infty &= \eps   \,
\end{aligned}
\ee
reproduces the  result \eqref{eq:HEEstatic}. 

The holographic entanglement entropy for the time dependent state is instead obtained with a different identification
\be
\begin{aligned} \label{eq:asympdiff}
w_\infty &= w_\tE (w = i \sigma) \\ 
\bar w_\infty & = \bar w_\tE (\bar w =- i \sigma)  \\ 
u_\infty &= \eps \sqrt{ w_\tE '(w = i\sigma) \bw_\tE '(\bw= - i\sigma)} \, .
\end{aligned}
\ee
This is the asymptotic form of the bulk diffeomorphism corresponding to the CFT map relating the initial state to the post quench state \cite{Banados:1998gg,Roberts:2012aq}.  
It pulls back the point $w_\infty,\bar w_\infty,u_\infty$ in the time independent geometry to the point $\sigma$ and bulk cutoff $\epsilon$ in the geometry dual to the time dependent state at time $t_E$. This can be though of as the bulk counterpart of the fact that on the CFT side an operator in the time dependent state at a given time $\tE$ can be expressed in terms of the initial static state operator through the map $w_\tE$.

Using the identifitcation  \eqref{eq:asympdiff} in \eqref{eq:Sholostatic}, and  continuing to Lorentzian time $\tE \to i t $,  it is  quite  immediate to see how this reproduces the post quench result. To  illustrate  this more explicitly we  consider the example of the SSD quench,  other cases follows with similar computations. Using in  \eqref{eq:asympdiff} the explicit form of the SSD time dependent map \eqref{eq:wtauSSD} and continuing to Lorentzian time $\tE \to i t $,  one can easily check that 
\be
u_\infty = \epsilon \sqrt{ w_t '(w = i\sigma) \bw_t '(\bw= - i\sigma)}= \epsilon \sqrt{ \frac{1}{  f(t)^2 +   \sin ^2   \(  \frac{2 \pi  \sigma}{L}  \)    }}
\ee
which matches the time depedent factor accompanying the cutoff in the expressions for the SSD entanglement entropy \eqref{eq:CFTresult}. Similarly,  in the argument of the $\sin$ in  the holographic formula \eqref{eq:Sholostatic} one recognizes  $\delta(t)$, the phase defined in \eqref{eq:CFTresult}
\be
 |\omega_t-\bar\omega_t| =  \frac{L}{\pi} \Big| \log \frac{z_t}{\bar z_t} \Big| =  \frac{L}{\pi} \delta(t)\, .
\ee
To be explicit  we used the relation between the CFT coordinates on the strip and in the UHP to write the middle expression.

\section{Discussion}\label{sec:conclusions}
 
We studied the role of boundary effects for the entanglement entropy in a CFT defined on an interval with mixed boundary conditions after an inhomogeneous quench. We  considered a quench protocol where the initial state is an eigenstate of  $L_0$ and the evolution is driven by a deformation of the standard CFT Hamiltonian. In particular, we analyzed the case of  the SSD and M\"obius deformation, as well as a general class of Hamiltonian deformations defined in terms of $sl(2,\mathbb{R})$ subalgebras of the full Virasoro algebra.  We  probed the entanglement dynamics in terms of the entanglement entropy for a portion of the system that includes one of the boundaries. 

Although the setup and the techniques employed are in large part  general, our analysis  focused on the case of holographic CFTs. In this context it is natural to focus on the universal contribution to the entanglement entropy coming from the identity Virasoro block. This is expected to  give the leading contribution at large $c$ and to capture the physics of semiclassical gravity in the holographic AdS$_3$ dual. 

Our analysis reveals an entanglement dynamics with a characteristic phase transition pattern governed by boundary effects. 
The mixed conformal boundary conditions,  $A$ and $B$,  in our setup give rise to two competing phases for the entanglement entropy obtained from  the conformal family of the stress tensor,  as described in section  \ref{sec:univSEE}. For the initial static state, one or the other phase dominates depending on the size of the entangling region and the relative value of the boundary entropies. 
The post-quench evolution, on the other hand,  turns out to allow for a dynamical transition between these phases. For a fixed entangling region, the transition pattern is  determined by the relative amount of  degrees of freedom associated with the two conformal  boundaries, as encoded in the boundary entropies $s_{A,B}$. When the number of degrees of freedom encoded in $s_A$ and $s_B$ are close enough to each other, in a sense made quantitive in section \ref{sec:Mobius}  and \ref{sec:GeneralH},   the quench dynamics induces a transition from the initial phase to the other. For boundary  entropies values that differ significantly, the initial phase instead dominates for all time. 
This is a common feature of all the deformations we considered in this work, but the details depend on the specific deformation considered. In particular the precise range of $s_B- s_A$ for which the transition occur depends on the deformation.    

At the kinematic level the entanglement dynamics of each  phase  is analogous to the case of  homogeneous boundary conditions \cite{Wen:2018vux}. For the  M\"obius and generalized $sl(2,\mathbb{R})$  deformations the entanglement  entropy exhibits a kinematics compatible with a finite size system, characterized by  an effective length which details are determined by the deformation parameters. This implies also a recurrent behaviour for the phase transition pattern. Within one period the phases alternate,  going from, \eg, phase $A$ to $B$, and back to the initial phase $A$, as depicted in fig.~\ref{fig:EEM} and \ref{fig:EEk}.
In the SSD case the kinematic is compatible with the one of an infinite dimensional system, with no return phase and a $\sim\log t$ growth of the entanglement entropy at lates times. Still, the presence of a phase transition signals the finiteness of the system with different conformal boundary conditions. 

As we discussed in section \ref{sec:holography}, our CFT analysis finds  a direct match on the geometric AdS$_3$  description. 
The holographic dual of the boundary CFT setup involves two EOW branes anchored on the boundary and extending in the AdS$_3$ bulk, where they intersect. This construction selects a portion of the AdS$_3$ geometry comprised between the  AdS$_3$ boundary and the two intersecting branes. In this setup the brane tensions  $T_A$ and $T_B$ are in one to one correspondence with the conformal boundary conditions $A$ and $B$ and determine the value of the boundary entropies.  
The two CFT channels described in section  \ref{sec:univSEE} have a direct correspondence  in the two competing  Ryu-Takayanagi surfaces one can draw in the asymptotically AdS$_3$ geometry. These are given by geodesics extending from the boundary of AdS to the EOW branes, as depicted in fig.~\ref{fig:interbranes}.

In section \ref{sec:HEEdyn}, we showed explicitly  how the holographic entanglement entropy exactly matches the CFT result. We obtained this result exploiting directly the local equivalence of asymptotically AdS$_3$ geometries and the dictionary relating CFT maps to bulk diffeomorphisms. 
We did not analyze explicitly the details of the  time dependent geometries  dual to the post quench state for the various deformations. It should be possible to do so along the lines of \cite{Goto:2021sqx,Goto:2023wai,Nozaki:2023fkx} which studied the case of a CFT on a circle.  

%
%
\paragraph{Double holography and quantum extremal surface\\}

Our analysis lends itself for an interpretation in the context of double holography. In fact, there is a third complementary description  which is available for our system, the so called brane perspective, as we schematically discuss in the following. 

There, the bulk spacetime enclosed between the asymptotic AdS boundary and the EOW branes is integrated out leaving behind an effective gravitational theory on each brane \cite{Randall:1999ee,Randall:1999vf,Karch:2000ct}. In the case of intersecting branes with a defect supporting the intersection,  we have two gravitational spacetimes, given by portions of AdS$_2$  with different effective cosmological constants. These are separated from one another by the defect on one side  and  in contact with the same non-gravitational CFT region on the other side. 

\begin{figure}[t!]
\centering
         \includegraphics[width=\textwidth]{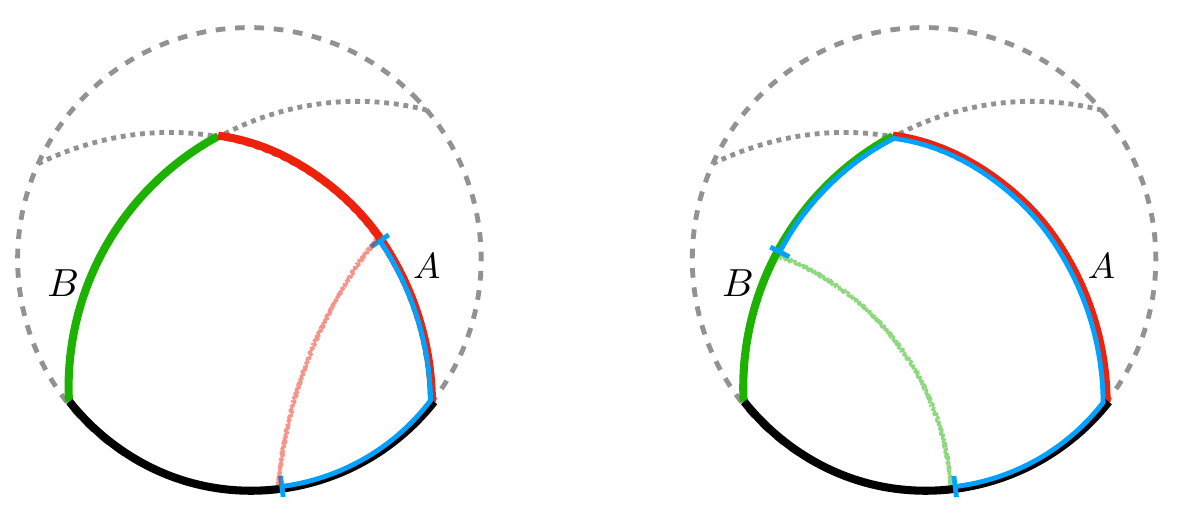}
\caption{A pictorial representation of the quantum extremal surface in the double holographic interpretation of our model. The RT surfaces of the AdS/BCFT setup select the QES in the brane perspective. The  phases $A$ and $B$ are depicted respectively on the left and on the right. In the $B$ phase the QES selects a region which comprises the entire AdS$_2$  region associated with  $A$ and part of the interior of the AdS$_2$ region associated with $B$.}
\label{fig:QES}
\end{figure}

Our entanglement entropy result can be interpreted in this perspective  as a generalized entropy computation \cite{Engelhardt:2014gca,Penington:2019npb,Almheiri:2019psf,Almheiri:2019hni,Almheiri:2019qdq,Almheiri:2020cfm}.  In the BCFT perspective, the region we considered includes the boundary with conformally invariant boundary condition $A$. In the brane description,  this corresponds to considering the entanglement entropy for  the interval $[0,\sigma]$ in the 2d CFT and an adjacent portion of the  gravitational space.  Which portion of the AdS$_2$ exactly enters is determined by the generalized entropy formula itself, which requires finding the quantum extremal surface (QES) \cite{Engelhardt:2014gca}. In double holography the location of the latter corresponds to the intersection of the Ryu-Takayanagi surface with the brane (see, \eg, \cite{Almheiri:2019hni,Geng:2020qvw,Chen:2020uac,Chen:2020hmv,Geng:2020fxl}).  The region  enclosed by the QES in the brane perspective then goes from $\sigma$ to this  point, as depicted in fig.~\ref{fig:QES}.

In the simplest case of a SSD quench, one can interpret the dynamical phase transition as follows:\footnote{Here we are thinking  in a quasi-static manner. A complete picture would require knowing the exact time dependent post quench geometry and the time dependent profile of the branes}  at the initial time of the post quench evolution the QES is inside universe $A$. As time evolves this goes deeper into the AdS$_2$ bulk of universe $A$. At the phase transition time, the QES surface jumps to universe $B$ selecting a region that extends from $\sigma$ all the way to cover the entire universe $A$, the bulk defect and part of the interior of universe $B$.

\paragraph{Driving the system with a sequence of Hamiltonians\\}

In our analysis we focused on quenching the system with one specific Hamiltonian. As discussed  for instance in \cite{Wen:2018agb,Wen:2020wee,Fan:2020orx,Wen:2021mlv,Lapierre:2019rwj,Lapierre:2020ftq,Jiang:2024hgt} for  the case of homogeneous boundary conditions, having an entire class of deformed Hamiltonian and a periodic behaviour allows for more complicated dynamical patterns. For finite $\theta$,   the system exhibits a periodic behaviour and gets revived to its initial state at the end of each period. One can then drive  the system with a different Hamiltonian at the end of each time period in a Floquet-like manner. 
The simplest case here  would be an evolution governed by two different Hamiltonians $H_{k_1}$ and $H_{k_2}$ at fixed $\theta$, with $n$-cycles
\be
U = \( e^{-i H_{k_2} t_2}  e^{-i H_{k_1} t_1}\)^n \, 
\ee
where $H_{k_i}$ drives the system for a period $t_i = \frac{2L\cosh(2\theta)}{k_i}$. 

Choosing appropriately the parameters of our setup, it would then be possible to engineer a variety of situations. For instance one could have a dynamics where a phase transition from phase $A$ to $B$ and back would happen for either, both or none the evolution intervals associated with $H_{k_1}$ and $H_{k_2}$. 

\paragraph{Generalizations\\}

Our explicit analysis  is limited to the case of holographic CFTs. Even though the complete set of conformal data is unknown, the holographic geometric construction in terms of EOW branes leads to consider arbitrary values for the boundary entropy. In turn this allows to always find a window of parameters where a transition occurs. In the case of rational CFTs, on the other hand, the boundary entropy can only take a finite number of values given by the $g$-functions \cite{Cardy:2004hm,Affleck:1991tk,Behrend:1998fd,Behrend:1999bn}. It would then be  interesting to revisit our analysis for the rational case and see if this dynamical phase transition is linked to the irrational character of  holographic CFTs, as it is the case for chaotic and scrambling behaviour (see, \eg, \cite{Roberts:2014ifa,Asplund:2015eha,Hosur:2015ylk}).

Also, we  only considered the case where the interval is adjacent to one of the boundaries. An interval which does not include any of the  boundaries will give rise to a more complicated pattern for the evolution of the entanglement entropy after the quench. An interesting observation that has been made for finite systems with periodic boundary conditions is that the SSD quench dynamics can be used to bring different initial states to a state where the entanglement entropy of an interval approximately matches that of the vacuum state at late times  \cite{Goto:2021sqx,Nozaki:2023fkx}.

Another interesting direction is extending the discussions to the thermal case. The relevant Euclidean CFT computation in this case is not on the infinite strip, but on the finite one with two sides identified, or equivalently on the annulus. Holographically it would correspond to considering an AdS black hole geometry with two branes selecting a part of the bulk. Despite a first principle CFT computation might be technically challenging in this case,  the same techniques of section~\ref{sec:holography} would allow to conduct an holographic study. We hope to report on this in the future.

%

\section*{Acknowledgement}

We thank Pawel Caputa, Yingfei Gu, Hosho Katsura,  Masamichi Miyaji, Yu Nakayama, Dominik Neuenfeld, Tatsuma Nishioka, Jiaxin Qiao, Erik Tonni and Satoshi Yamaguchi for useful discussions and comments at various stages of this work, and especially Tadashi Takayanagi  for comments on a draft of the paper. 
The authors are grateful for the hospitality of  Galileo Galilei Institute for Theoretical Physics (GGI).
 AB and FG are grateful for the hospitality of Perimeter Institute where part of this work was carried out.  
DG thanks the University of Florence and INFN for partial support during the completion of this work, and  IAS Tsinghua and YITP Kyoto where the work was partially presented. DG also thanks YITP for providing regular visiting opportunities.  DG is supported by  Grant-in-Aid for Transformative Research Areas(A)  ``Extreme Universe''  No. 21H05190. 
FG would like to thank the Isaac Newton Institute for Mathematical Sciences, Cambridge, for support and hospitality during the programme Black holes: bridges between number theory and holographic quantum information where work on this paper was undertaken. 
This research was supported in part by the Simons Foundation through the Simons Foundation Emmy Noether Fellows Program at Perimeter Institute. Research at Perimeter Institute is supported by the Government of Canada through the Department of Innovation, Science and Economic Development and by the Province of Ontario through the Ministry of Research, Innovation and Science.
This work was also supported in part by EPSRC grant no EP/R014604/1.
 

\appendix

\section{Stress tensor in the UHP}\label{App:MIM} 
 
On the infinite strip of width $L$  with conformal boundary conditions on the edges   only one set of Virasoro symmetries   survives. To see that, one can use the conformal map
\be\label{eq:striptoUHP}
z = e^{\frac{\pi}{L}w}\,,~~~ w=\tau +i \sigma\,,~~~\sigma \in [0,L]
\ee
to bring the strip to the UHP. See, \eg, \cite{Cardy:2004hm,DiFrancesco:1997nk}.

In the UHP Virasoro generators are still defined in terms of the stress energy tensor, though its holomorphic part and anti-holomorphic part are no longer independent from each other. More precisely, they conformal boundary conditions imply  $T(z) = \bar T(\bz)$ when $z=\bar z$, \ie, on the real axis. One can  analytically continue the holomorphic stress tensor to the lower half plane defining  $T(z^*) = \bar T (z)$, where $z$ in the UHP. Similarly for the anti-holomorphic part of the stress tensor $\bar T(z^*) = T (z)$.  

Then the Virasoro generators can be expressed in the UHP in a similar way as in the full complex plane 
\be\label{eq:ViraUHP}
L_n = \frac{1}{2\pi i}\oint_{C'} dz z^{n+1} T(z)  - \frac{1}{2\pi i}\oint_{C'} dz^*  {z ^*}^{n+1}\bar T(z)
\ee
where the contour $C'$  is a semicircle and a segment along the real axis including the origin. Using the analytic properties of the stress tensor this can also be expressed as a contour integral in the full complex plane
\be
L_n = \frac{1}{2\pi i} \oint_C dz z^{n+1} T(z) = - \frac{1}{2\pi i} \oint_C d\bar z \bar z^{n+1} \bar T(\bar z)= \bar L_n\,,
\ee
In the main text we will express the Virasoro generators as contour  integrals in the full complex plane using the holomorphic part and use the short-hand notation 
\be
 \oint z^{n+1} T(z)    \equiv \frac{1}{2\pi i} \oint_C dz z^{n+1} T(z)  = L_n \,. 
\ee
Notice that $L_n$ now acts both on the holomorphic and anti-holomorphic coordinates of the bulk primary operator as
\be\label{eq:ViraOcm}
[L_n, \mO(z,\bar z)] = h(n+1) z^{n}\mO(z, \bar z) + z^{n+1}\der_z \mO(z,\bar z) + \bar h(n+1) \bar {z}^{n}\mO(z, \bar z) + \bar z^{n+1}\der_{\bar z} \mO(z,\bar z)\, .
\ee
%



\end{document}